\documentclass[10pt,journal,compsoc]{IEEEtran}

\IEEEoverridecommandlockouts       
\usepackage{pifont}
\usepackage{bbm}
\usepackage{xcolor}
\usepackage[utf8x]{inputenc}
\usepackage{fancybox}
\usepackage{verbatim}
\usepackage{graphics}
\usepackage{smartdiagram}
\usepackage{scrextend}
\usepackage{hyperref}
\usepackage{multicol}
\usepackage[T1]{fontenc}
\usepackage[english]{babel}
\usepackage[linecolor=white]{mdframed}
\usepackage{subcaption}
\usepackage[compatibility = false]{caption}
\usepackage{overpic}
\usepackage{comment}
\usepackage{multirow} 
\usepackage{graphicx}
\usepackage{eurosym}
\usepackage{booktabs}
\usepackage{placeins}
\usepackage{algorithmicx}
\usepackage{algpseudocode}
\usepackage{algorithm}
\usepackage{adjustbox}
\usepackage{enumerate}
\usepackage{paralist}
 \usepackage{siunitx}
 \usepackage{amsmath,amssymb,graphics}
\usepackage{bbold}
 \usepackage{circuitikz}
 \usepackage{tikz}
 \usetikzlibrary{shapes.geometric}
 \usetikzlibrary{automata}
 \usetikzlibrary{positioning}
 \usetikzlibrary{calc}
\usepackage{empheq}
 \usepackage{cite}
 \usepackage{pgfplots}
 \usepackage{setspace}
 \usepackage{soul,color,cancel}
 \tikzstyle{block} = [draw, rectangle, minimum width=1cm, minimum
 height =0.75cm]
 \tikzstyle{sum} = [draw, circle, node distance=0.75cm]
 \tikzstyle{input} = [coordinate] \tikzstyle{output} = [coordinate]
 \tikzstyle{pinstyle} = [pin edge={to-,thin,black}]
\pgfplotsset{
scaled y ticks = false,
scaled x ticks = false,
grid=major,
xtick={0,7200,14400,21600},
xticklabels={{0},{120},{240},{360}},
grid style={dashed,gray!70,very thin},
width=4.3cm,
height=4cm,
}

\newcommand{\cmark}{\ding{51}}%
\newcommand{\xmark}{\ding{55}}%
\newcommand{\R}{\mathbb{R}}

\newcommand{\N}{\mathbb{N}}

{\left(\begin{smallmatrix}}
{\end{smallmatrix}\right)}

\title{Imitation and Heterogeneity Shape the 
Resilience of Community Currency Networks}

\author{C. Ancona$^{\dag}$, D. Ricci$^{\natural}$, C. Bernardo$^{\natural}$, F. Lo Iudice$^{\dag}$, A. Proskurnikov$^{*}$, F. Vasca$^{\natural}$
\thanks{$^{\dag}$ University of Naples Federico II,  
emails:  {\tt camilla.ancona2@unina.it}, {\tt francesco.loiudice2@unina.it}}
\thanks{$^\natural$ Department of Engineering, University of Sannio, Be\-ne\-ven\-to, Italy, emails: {\tt dora.ricci@unisannio.it}, {\tt carmela.bernardo@unisannio.it}, {\tt vasca@unisannio.it}}
\thanks{$^{*}$ Department of Electronics and Telecommunications, Politecnico di Torino, email: {\tt anton.p.1982@ieee.org}}
\thanks{The work was supported by the project 2022K8EZBW ``Higher Order Interactions in Social Dynamics with Application to Monetary Networks'' funded by European Union-Next Generation EU within the PRIN 2022 Program (D.D. 104, 02/02/2022, the Ministry of University and Research). This manuscript reflects only the authors’ views and opinions, and the Ministry cannot be considered responsible for them.}
}

\pgfplotsset{compat=1.18} 
\begin{document}
\maketitle
\thispagestyle{empty}
\pagestyle{empty}


\begin{abstract}
Community currency networks are made up of individuals and/or companies that share some physical or social characteristics and engage in economic transactions using a virtual currency. This paper investigates the structural and dynamic properties of such mutual credit systems through a case study of Sardex, a community currency initiated and mainly operating in Sardinia, Italy. The transaction network is modeled as a directed  weighted graph and analyzed through a graph-theoretic framework focused  on the analysis of strongly connected components, condensed representations, and behavioral connectivity patterns. Emphasis is placed on understanding the evolution of the network’s core and peripheral structures over a three-year period, with attention to temporal contraction, flow asymmetries, and structural fragmentation depending on different user types. Our findings reveal persistent deviations from degree-based null models and suggest the presence of behavioral imitation, specifically, a user preference for more active peers. We further assess the impact of heterogeneous connections between different type of users, which strengthen the network topology 
and enhance its resilience. 
\end{abstract}
\section{Introduction}

Complementary currencies are means of exchange designed to supplement national currencies, frequently aimed at strengthening local economies or achieving specific social objectives. A \textit{community currency} (CC) represents a specialized form of complementary currency, used primarily within clearly defined communities, which may include geographically localized groups, business networks, or digital communities \cite{ijccr_ws}. CCs typically operate within a \textit{local exchange trading system}, a community-based framework where transactions are democratically governed, not-for-profit, and based on trust among members exchanging goods and services using locally generated currency.

One prominent example of such a system is \textit{Sardex}, a mutual credit CC initiated in 2009 in Sardinia, Italy, that primarily supports business-to-business transactions but has evolved to facilitate partial salary payments \cite{sardex_ws}. Sardex’s success largely hinges on mutual trust and reciprocal relationships among participants, highlighting the importance of community structures.

Research studies on the Sardex mutual credit system can be broadly categorized into two main groups: (a) those focusing on its social, institutional, and political dynamics and its socio-economic perspectives, and (b) those that apply quantitative methodologies from network science. Among the earliest contributions in the first line of research is the paper by Sartori and Dini~\cite{sartori2016}, which offers a micro–macro perspective on Sardex's emergence as a local institution and its role in fostering trust-based economic relations within Sardinia. Motta et al. \cite{motta2017} use 29 semi-structured interviews to argue that Sardex functions as a form of self-funded social impact investment, integrating market activity with democratic institutions and cultural values. Littera et al.  \cite{littera2017} frame Sardex as a social innovation startup that carefully balances economic benefits with social cohesion, highlighting the system's foundation in trust. The political significance of Sardex is explored by Kioupkiolis and Dini \cite{kioupkiolis2019}, describing it as a space for collective micropolitical engagement and a practical alternative to dominant economic paradigms. Bazzani \cite{bazzani2020} explores the role of money in modern society, comparing the traditional model with Sardex. 

In this work, however, we pursue the line of research which employs network-based and other quantitative analytical frameworks to study CC systems. These systems exhibit an inherently networked nature, in which participants and transactions naturally form complex relational structures. These analyses typically represent CC exchanges through directed weighted graphs where nodes signify users and edges indicate monetary transactions. Prior research has utilized network science to study CCs across diverse contexts, including:
\begin{itemize}
\item \textit{Sardex}, Italy, established in 2009 \cite{iosifidis2018};
\item \textit{Sarafu}, Kenya, launched in 2018 \cite{mattsson2022, mqamelo2022community, Mattsson2023Circulation};
\item \textit{Ichi-Muraoka}, Japan, introduced in 2002 \cite{nakazato2012};
\item \textit{Tomamae-cho}, Japan, active briefly in 2004--2005 \cite{Kichiji2008Network};
\item \textit{Peanuts}, Japan, established in 1999 \cite{nakazato2015};
\item \textit{Wymiennik-ALTERKA}, Poland, from 2012 \cite{lopaciuk2019};
\item \textit{RotsLETSe}, Czech Republic, active since 1999 \cite{frankova2014};
\item \textit{Bytesring Stockholm}, Sweden, launched in 1992 \cite{nakazato2012}.
\end{itemize}

Despite the considerable body of research, key structural and temporal aspects of CCs remain insufficiently explored, particularly regarding the evolution and interconnectivity of network components over time and their implications for liquidity circulation and economic resilience \cite{ba2023cooperative, nakazato2024multiplex}. 

In the remaining part of this section, the literature related to our work is discussed, and our contributions are highlighted. The rest of the paper is structured as follows: Section~\ref{sec:data} presents our dataset, Section~\ref{sec:metrics} shows a preliminary analysis based on the metrics of its networks, Section~\ref{sec:cond} details the structural analysis results, Section~\ref{sec:nm} compares the Sardex network with a network null model to unveil the imitation strategy that users utilize, Section~\ref{sec:multi} unravels the key role of users heterogeneity in the resilience of the network, and Section~\ref{sec:conc} concludes with discussions and future research directions.

\subsection{Literature on network analysis for CCs}
\label{sec:lr}

Several distinct research directions are prominent in the analysis of CC networks from a network science perspective (see Table~\ref{table:comp} for a list of recent studies). 

One research direction focuses on the detection of community structures within the CC circuit.
The homophily (heterophily) concept in terms of node degree is used in~\cite{Kichiji2008Network} to derive assortative (disassortative) mixing, i.e., the tendency of high-degree vertices to attach to other high-degree (low-degree) vertices. Analogously, for community detection, the “rich-club” coefficient is considered in~\cite{frankova2014}, and the research dependence is analyzed in~\cite{nakazato2015}. Map equations and the associated Infomap algorithm are used in~\cite{Mattsson2023Circulation, bohlin2014} to study circulation. Through this algorithm, one obtains a hierarchical clusterization of nodes which are grouped in terms of intensity of flow observed between them (and little outside). The composition of these subpopulations can then be understood using an approach in which their heterogeneity is quantified with respect to node attributes \cite{kelly2014, kelly2012}. 
Broader digital CC design principles and classifications have been explored in \cite{Chasin2020Design, Diniz2018Taxonomy}. Several studies have leveraged blockchain technology, enabling precise and transparent transaction tracking. Blockchain applications for local complementary currencies have been addressed in \cite{Dominique2019Proof}. Mqamelo \cite{mqamelo2022community} has provided a pioneering randomized controlled trial demonstrating significant economic impacts of blockchain-based CC transfers. Similarly, Ba et al. \cite{ba2023cooperative} have analyzed cooperative behaviors within blockchain-based CC networks during crises, emphasizing temporal and geographic influences.

The second direction of research focuses on quantifying CC graph representations using centrality metrics, such as reciprocity, cycles, clustering coefficients, eigenvector centrality, PageRank, and transitivity 
\cite{Kichiji2008Network, Katai2009Fuzzy, nakazato2015, lopaciuk2019, iosifidis2018, frankova2014, Mattsson2023Circulation}. Reciprocity captures users’ tendencies to form reciprocal exchange relationships central to CC systems \cite{Kichiji2008Network, Katai2009Fuzzy}. Cycles, indicating liquidity circulation, are critical in sustaining economic activity within CCs \cite{iosifidis2018, Mattsson2023Circulation}. Network metrics like clustering coefficients and PageRank provide insights into community cohesiveness and node prominence, respectively \cite{frankova2014, Kichiji2008Network, Mattsson2023Circulation}. Studies such as Iosifidis et al. \cite{iosifidis2018} and Mattsson et al. \cite{Mattsson2023Circulation} have particularly emphasized cyclic motifs and modular network structures, highlighting the role of cycles in enhancing economic robustness. Mattsson et al. have also utilized the Infomap algorithm for hierarchical clustering, which effectively identifies community structures based on transactional flow intensity.

The literature analysis discussed above is summarized and categorized in Table~\ref{table:comp}. 
This comparative framework highlights the novelty of our approach (last line in the table), which integrates multiple network-theoretic dimensions that are absent from most existing works.

\begin{table}[!h]
\centering
\resizebox{\columnwidth}{!}{
\begin{tabular}{l l c c c c c c c c}
\toprule
\textbf{Paper} & \textbf{Dataset} & \textbf{Weights} & \textbf{Recip./Cycles} & \textbf{Condensed} & \textbf{Null Model} & \textbf{Geo} & \textbf{Temporal} & \textbf{Clustering} & \textbf{Blockchain} \
\\ \midrule
\vspace{3pt}
\cite{iosifidis2018} & Sardex & \cmark & \cmark & \xmark & \cmark & \cmark & \xmark & \xmark & \xmark \\
\vspace{3pt}
\cite{mqamelo2022community} & Sarafu & \cmark & \xmark & \xmark & \xmark & \cmark & \cmark & \xmark & \cmark \\
\vspace{3pt}
\cite{Mattsson2023Circulation} & Sarafu & \cmark & \cmark & \xmark & \xmark & \cmark & \cmark & \cmark & \cmark \\
\vspace{3pt}
\cite{Kichiji2008Network} & Tomamae-cho & \cmark & \cmark & \xmark & \xmark & \cmark & \xmark & \xmark & \xmark \\
\vspace{3pt}
\cite{nakazato2015} & Peanuts & \xmark & \cmark & \xmark & \xmark & \xmark & \xmark & \xmark & \xmark \\ 
\vspace{3pt}
\cite{lopaciuk2019} & ALTERKA & \xmark & \cmark & \xmark & \xmark & \cmark & \xmark & \xmark & \xmark \\
\vspace{3pt}
\cite{frankova2014} & RozLEŤSe & \cmark & \xmark & \cmark & \xmark & \xmark & \xmark & \xmark & \xmark \\
\vspace{3pt}
\cite{ba2023cooperative}& Sarafu & \cmark & \xmark & \xmark & \xmark & \cmark & \cmark & \cmark & \cmark \\
\vspace{3pt}
\cite{nakazato2024multiplex} & Hanbat LETS  & \cmark & \cmark & \xmark & \cmark & \xmark & \xmark & \cmark & \xmark \\
\vspace{3pt}
\cite{bohlin2014} & \xmark & \xmark & \xmark & \xmark & \cmark & \xmark & \xmark & \cmark & \xmark \\
\vspace{3pt}
\cite{kelly2014} & \xmark & \cmark & \cmark & \xmark & \xmark & \cmark & \cmark & \cmark & \xmark \\
\vspace{3pt}
\cite{kelly2012} & \xmark & \cmark & \cmark & \xmark & \cmark & \cmark & \cmark & \cmark & \xmark \\
\vspace{3pt}
\cite{Chasin2020Design} & \xmark & \xmark & \xmark & \xmark & \xmark & \xmark & \xmark & \cmark & \cmark \\
\vspace{3pt}
\cite{Diniz2018Taxonomy} & \xmark & \cmark & \cmark & \xmark & \xmark & \cmark & \cmark & \cmark & \cmark \\
\vspace{3pt}
\cite{Dominique2019Proof} & \xmark & \xmark & \xmark & \xmark & \cmark & \xmark & \xmark & \xmark & \cmark \\
\vspace{3pt}
\cite{Katai2009Fuzzy} & \xmark & \xmark & \cmark & \xmark & \xmark & \xmark & \xmark & \cmark & \xmark \\
\vspace{3pt}
\cite{dorogovtsev2001giant} & \xmark & \xmark & \xmark & \cmark & \xmark & \xmark & \xmark & \xmark & \xmark \\
\vspace{3pt}
\cite{criscione2025topological} & Sarafu & \cmark & \cmark & \cmark & \cmark & \cmark & \cmark & \xmark & \xmark \\
\vspace{3pt}
\textbf{This paper} & Sardex  & \cmark & \cmark & \cmark & \cmark & \cmark & \cmark & \cmark & \xmark \\
\\ \bottomrule
\end{tabular}
}
\caption{Comparison of selected studies analyzing CCs from the network science and engineering viewpoint. 
Each row represents a different study. Columns 3-10 indicate whether the study: (1) uses transaction volumes as weights; (2) examines reciprocity and/or cycles; (3) applies component or condensation graph analysis; (4) includes a null model comparison; (5) considers geographic dimensions of transaction patterns; (6) incorporates a temporal analysis of network evolution; (7) employs clustering or community detection methods; and (8) investigates blockchain infrastructure for currency implementation or analysis.}
\label{table:comp}
\end{table}

\subsection{Contributions}

This work makes several contributions to the analysis of CC networks by applying advanced graph theory tools and community detection techniques to the real-credit Italian network Sardex. Through a combination of topological, temporal, geographic, and behavioral analyses, the study 
shows how behavioral imitation and user-type heterogeneity shape the structural resilience of CC networks. 
%
While our methods draw from established tools in graph theory and statistical mechanics, their integrated and extended application across multiple years and comparison with null models allows us to show that imitation and heterogeneity are two key factors determining the prosperity and resilience of CCs.


The first key contribution of this paper consists of showing that currency circulation in CC networks is boosted by \textit{imitation}, i.e., a strategy‐updating rule where agents revise their behavior by observing and copying more successful neighbors \cite{wang2023imitation,bara2022enabling}, which in our context is interpreted as the tendency of less active members to engage in transactions with nodes characterized by higher activity levels, giving less attention to their personal information.
Our results show that CC users transact with peers whose out-degree is relatively homogeneous, and asymmetry indices reveal a statistically significant bias toward transacting with more active peers. This suggests a prestige-like preference or imitation dynamic not previously identified in CC network studies, which can be interpreted as a peculiar type of preferential attachment.

Second, we demonstrate that \textit{heterogeneity} represents a further key factor for the CC network resilience. This is done through a comprehensive analysis of condensation graphs for a CC network—previously partially used only in \cite{criscione2025topological} for Sarafu— in a three-years temporal framework, while prior studies such as \cite{iosifidis2018} and \cite{Mattsson2023Circulation} have captured cyclicity in single-year data. The bow-tie network decomposition  \cite{dorogovtsev2001giant}  has provided a mesoscopic lens on the transactional architecture and complements traditional metrics like reciprocity or clustering. By weighting these condensed components not just by node count but also by transaction volume and frequency, we reveal a consistent contraction of the giant strongly connected component (GSCC) over time accompanied by a proportional expansion of its downstream. This dynamic reconfiguration suggests weakening in recirculating monetary flow and increasing structural dependence on sink nodes.
%
%
%
Moreover, the multilayer representation allows us to identify the role of user-type heterogeneity in sustaining network connectivity. By grouping users as businesses or persons, we show that inter-type (interlayer) connections are disproportionately responsible for expanding the GSCC. These heterogeneous ties act as structural bridges across community subgroups, reinforcing the core. This multi-layered perspective is absent from prior CC studies and provides new insight into how participant diversity enhances system resilience.


In synthesis, this study brings together insights from multiple levels of network structure while bridging quantitative modeling with socio-economic interpretations. By contextualizing Sardex within a broader ecosystem of CCs and highlighting previously unexamined structural behaviors, we offer new foundations for both theoretical investigation and practical design of resilient CC networks.

\section{Dataset and network construction}
\label{sec:data}
This section describes the dataset used for the analysis, consisting of all Sardex transactions recorded in the time frame from January 2022 to December 2024 and describes the construction of the three graphs based on corresponding annual data (2022, 2023, 2024).

\subsection{Notation}

The currency circulation is determined by transactions between different users. Each transaction is identified by its amount, a date, a seller, and a buyer. The rules of Sardex impose that 1 Sardex is equivalent to 1 Euro; for the sake of simplicity, the symbol \euro~will be used throughout the paper for indicating the volume of Sardex transactions.

In the following, $\N$ ($\R$) is the set of natural (real) numbers, $\R^+$ the set of nonnegative real numbers, $|\mathcal{N}|$ indicates the size of the set $\mathcal{N}$. 
The dataset is analyzed by using the digraph $\mathcal{G} = \{\mathcal{N}, \mathcal{E}\}$, where $\mathcal{N}=\{1, \dots, N\}$, $N \in \N$, denotes the set of nodes which are the CC  users and $\mathcal{E} \subseteq \mathcal{N} \times \mathcal{N}$ is the set of edges. An edge from node $i \in \mathcal{N}$ to node $j\in \mathcal{N}$ exists if there was at least one transaction from $i$ (the buyer) to $j$ (the seller) during the year. The binary variable 
$\delta_{ij} \in \{0,1\}$ indicates whether an edge from $i$ to $j$ exists ($\delta_{ij}=1$) or not ($\delta_{ij}=0$). It is assumed that $\delta_{ii}=0$ for all $i \in \mathcal{N}$, i.e., self-loops are disregarded. The set of (out-)neighbors of the $i$-th node is defined as $\mathcal{N}_i=\{j \in \mathcal{N} : \delta_{ij}=1\}$, $i \in \mathcal{N}$.  
The two standard weightings for the edges are: the \textit{number of transactions} $e_{ij} \in \N$, representing the total number of outgoing transactions from $i$ to $j$, and the  \textit{total volume} $w_{ij} \in \R^+$, standing for the total amount transferred from $i$ to $j$ over the year.
The \textit{in-degree} (\textit{out-degree}) of the $i$-th node, say $\theta^{\text{in}}_i = \sum_{j \in \mathcal{N}} e_{ji}$ ($\theta^{\text{out}}_i = \sum_{j \in \mathcal{N}} e_{ij}$), is the total number of incoming (outgoing) transactions for the corresponding user over the year. 
For the $i$-th node, the total \textit{incoming volume} (\textit{outgoing volume}) over the year is given by $v_i^{\text{in}}=\sum_{j \in \mathcal{N}} w_{ji}$ ($v_i^{\text{out}}=\sum_{j \in \mathcal{N}} w_{ij}$).  
For any two nodes $i, j \in \mathcal{N}$, $j$ is \textit{reachable} from $i$ if there exists a directed path from $i$ to $j$. Given a subset of nodes $\mathcal{S} \subset \mathcal{N}$, the \textit{downstream set} $\mathcal{D}(\mathcal{S})$ includes all nodes that can be reached from any node in $\mathcal{S}$. Conversely, the \textit{upstream set} $\mathcal{U}(\mathcal{S})$ consists of all nodes from which any node in $\mathcal{S}$ can be reached. A \textit{strongly connected component} (SCC) of a digraph $\mathcal{G}$ is a maximal subgraph where every node is reachable from every other node within the subgraph. Any digraph $\mathcal{G}$ can be decomposed into a finite number of SCCs.
%
A \textit{weakly connected component} of a digraph $\mathcal{G}$ is a subgraph of $\mathcal{G}$ whose vertices are connected to each other by a path that can be constructed ignoring the direction of its edges. An \textit{acyclic graph} is a graph that contains no cycles, i.e., no closed paths from any node back to itself. Given a sequence of vectors $\{\xi_i\}_{i \in \mathcal{N}}$, we use symbol $\xi=\text{col}(\{\xi_i\}_{i \in \mathcal{N}})$ to denote 
the vector obtained by stacking the entries $\xi_i$, $i \in \mathcal{N}$, into a single column. 

\subsection{Dataset description}



The Sardex network consists of four different types of users: business (B), consumer (C), employee (E), and provider (P). Other information available for each user are sector, activity, and province. The number of users and volumes in the three years of interest, partitioned by type of user, are reported in  Table~\ref{tab:typememb}. A comparison with the data presented in~\cite{iosifidis2018} for the years 2013 and 2014 shows that the Sardex circuit has grown by an order of magnitude in the past decade. However, the permanence of many users inside the circuit remains volatile: about one-third of nodes active in 2022 had exited the market by 2023, with this percentage rising to 60\% for nodes from 2023 that exited the market by 2024. On the other hand, users who left the circuit were involved only in low volumes of transactions (see Appendix~\ref{app1}). In general, users who left the circuit were mostly type C, while those who remained were primarily B users.

Additional considerations regarding the role of different type of users can be derived by considering the volumes of transactions. Observing the values in Table~\ref{tab:typememb}, the majority of exchanged volumes, exceeding 84\% of the total in all three reference periods, can be attributed to B users, although C users represent the largest user group in each period, except for 2022, when the numbers were nearly identical. Analyzing the transaction distribution by user types of buyers and sellers (see Appendix~\ref{app2}) confirms  a strong concentration of economic activities around B users. The latter represents the network's transactional core, handling approximately 70\% of purchases and over 80\% of the total traded volume. On the sales side, an equally significant pattern can be observed: while all types of users actively participate in transactions, the relevant volumes are recorded almost exclusively in interactions with B users. In other words, regardless of the category they belong to, users tend to sell mainly to this type, consolidating their central role in the market.

Even though the analysis above could suggest a dominant position or a structural preference towards B users, the other type of users also play a fundamental role for the circuit resilience so as the condensed graph and multilayer network analyses will show in the next sections. 

\begin{table}[!h]
    \centering
    \resizebox{0.48\textwidth}{!}{
    \begin{tabular}{rrrrrrr} 
    \hline
    Year&$t$&\multicolumn{1}{c}{B}&\multicolumn{1}{c}{C}&\multicolumn{1}{c}{E}&\multicolumn{1}{c}{P} &\multicolumn{1}{c}{Total}\\ \hline
    \multirow{2}{*}{2022}&$N_t$&5,461&6,604&2,581&3&14,649\\
    &$v^{\text{out}}_{i \in \mathcal{N}, {i = t}}$&51,302\,k\euro&354\,k\euro&6,647\,k\euro&1,889\,k\euro&60,193\,k\euro\\\hline\hline
    Year&$t$&\multicolumn{1}{c}{B}&\multicolumn{1}{c}{C}&\multicolumn{1}{c}{E}&\multicolumn{1}{c}{P} &\multicolumn{1}{c}{Total}\\ \hline
    \multirow{2}{*}{2023}&$N_t$&5,343&14,452&2,858&4&22,657\\
    &$v^{\text{out}}_{i \in \mathcal{N},{i = t}}$&52,293\,k\euro&338\,k\euro&6,814\,k\euro&2,224\,k\euro&61,668\,k\euro\\\hline \hline
    Year&$t$&\multicolumn{1}{c}{B}&\multicolumn{1}{c}{C}&\multicolumn{1}{c}{E}&\multicolumn{1}{c}{P} &\multicolumn{1}{c}{Total}\\ \hline
    \multirow{2}{*}{2024}&$N_t$&4,744&6,170&2,788&2&13,704\\
    &$v^{\text{out}}_{i \in \mathcal{N}, {i = t}}$&47,215\,k\euro&177\,k\euro&6,335\,k\euro&2,018\,k\euro&55,745\,k\euro\\ \hline
    \end{tabular}}
\caption{Number of users $N_t$ and outgoing volumes of Sardex for different  years and type $t \in \{$B,C,E,P$\}$.}
\label{tab:typememb}
\end{table}

\subsection{Different ranges of transactions amount}

\begin{figure}[ht]
\centering
\includegraphics[width=0.8\linewidth]{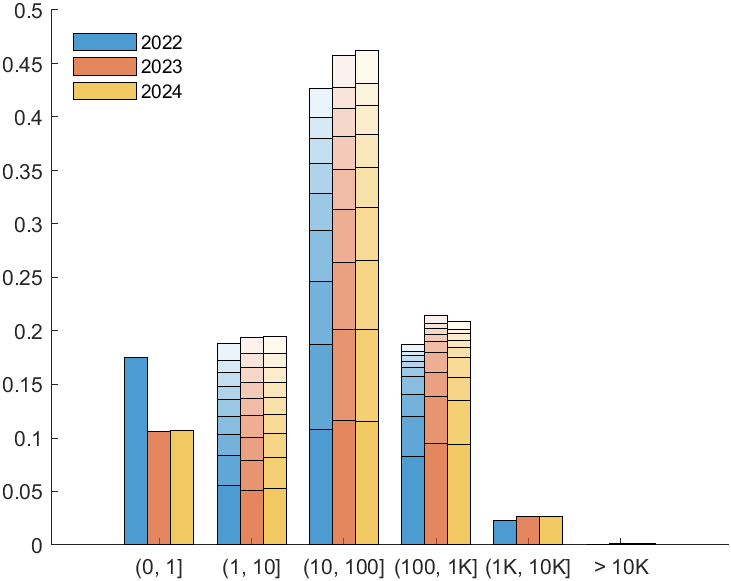}
\caption{Distribution of the number of transactions (with respect to the total number of transactions) for different ranges of monetary values. Inside each vertical bar of the central ranges in the figure, the sub-intervals are at steps of $10\%$ of the corresponding maximum value.}
\label{fig:barreimpilate}
\end{figure}

The circuit allows transactions of any amount. Figure~\ref{fig:barreimpilate} shows the relative number of transactions for different ranges of monetary values. 

The distribution of transactions by user type (see Table~\ref{tab:tabVolumi} in Appendix~\ref{app2}) highlights that approximately 60\% of transactions carried out by B users involve amounts between 10 and 100\euro. For C users, more than 50\% of transactions occur with amounts below 1\euro. Over 65\% of transactions by E users fall within the range of 100 to 1,000\euro. As for P users, transaction patterns vary across years: in 2022 (65.1\%) and 2024 (37.0\%), the most frequent transactions range between 100 and 1,000\euro, whereas in 2023 (42.6\%), the majority falls within 1 to 10\euro.

\begin{figure}
\centering    \includegraphics[width=0.8\linewidth]{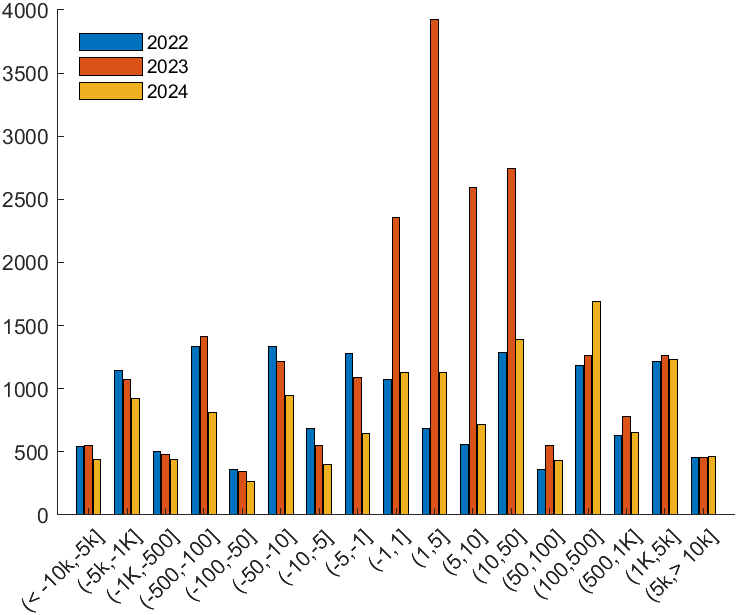}
\caption{Distribution of yearly net balance of users.}
\label{fig:netbal}
\end{figure}

Each user can sell and buy. The distribution of the number of users by net volume over the years is shown in Figure~\ref{fig:netbal}. A significant portion of users has a nearly balanced account, falling within the range $(-5,5]$\euro. Expanding the analysis to the range  $(-50,50]$\euro, the percentage of users within this bracket varies from $46\%$ in 2024 to $64\%$ in 2023. However, the distribution of balances is asymmetric, with the share of users holding a positive balance ranging from $45\%$ in 2022 to $68\%$ in 2023.
A closer examination of different user groups (see Table~\ref{tab:tabBilanci} in Appendix~\ref{app2}) reveals that:
\begin{itemize}
    \item most B users have balances beyond the range $(-50,50]$\euro, and the B users with positive balance dominate;    
    \item over $70\%$ of C users have balances within the $(-50,50]$\euro~range;
    \item for E users, however, the percentage of those within this range drops to approximately $25\%$.
\end{itemize}

\section{Network centrality metrics} \label{sec:metrics}

A preliminary study of a CC network can be performed by analyzing the centrality metrics of the corresponding graphs, as shown in this section by considering the case of Sardex.

\subsection{Degree-volume correlation}

\begin{figure}
    \centering
     \includegraphics[width=0.8\linewidth]{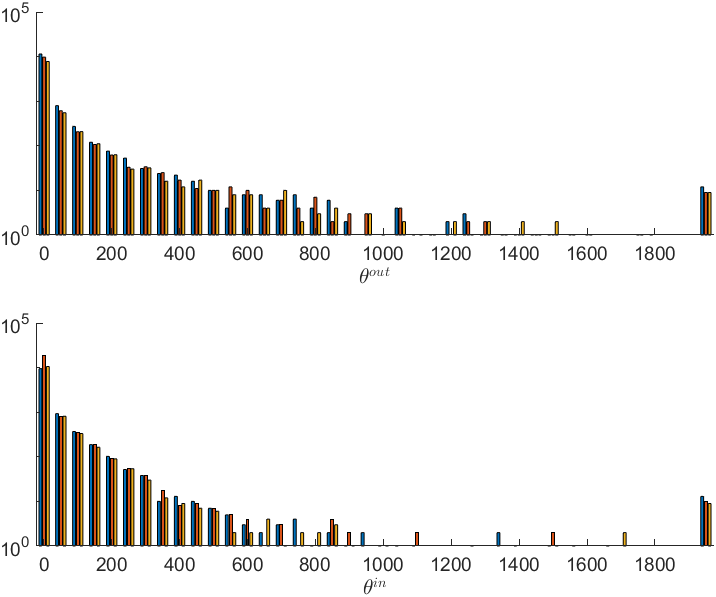}
    \caption{Distribution of degrees for the three data sets 2022, 2023, and 2024  (vertical axis scale is set to logarithmic scale, nodes with a degree higher than 2,000 have been merged into the last group of bars).}
\label{fig:deg_dist_comparison}
\end{figure}

The degree distributions remain consistent across all the periods analyzed, as shown in Figure \ref{fig:deg_dist_comparison}. The peak of these distributions occurs at lowest degree values, while the trend gradually decreases for  higher values of the degrees. Users with more than one thousand of  transactions, i.e., number of $i \in \mathcal{N}$ such that $\theta^{\text{in}}+\theta^{\text{out}}\ge 1000$, are $48$ in $2022$, $44$ in $2023$, and $42$ in $2024$. The average volume per user is $4$\,k\euro~in $2022$ and $2024$ and $3$\,k\euro~in $2023$. 

Users with high connectivity play a crucial role:
\begin{itemize}
    \item Those with an out-degree above $40$ constitute a small but influential group, responsible for over $75\%$ of outgoing transactions, with approximately half of the total outgoing transaction volume.
    \item Similarly, users with an in-degree above $40$ contribute to around $80\%$ of incoming transactions, corresponding to nearly two-thirds of the total incoming transaction volume.
\end{itemize}

A more detailed analysis of transactions and corresponding volumes (see Table~\ref{tab:i_numeri} in Appendix~\ref{app1}) highlights some important features:
\begin{itemize}
    \item The reduction in edges when considering binary connections is significant, dropping by approximately $70\%$ across all years. This effect is even more pronounced for nodes where both $\theta^{\text{in}}$ and $\theta^{\text{out}}$ exceed one.
    \item On average, $95\%$ of the total transaction volume comes from users who engage in both incoming and outgoing transactions, representing a substantial portion of the network.
    \item Nearly $90\%$ of the total volume originates from users who perform at least two transactions in both directions.
\end{itemize}

It is interesting to analyze the correlation between transactions and volumes. 
Table~\ref{tab:correlazione} shows that the most correlated variables ($0.93$ in 2022 and 2023, and $0.89$ in 2024) are the outgoing and incoming volumes,  showing the satisfaction of the balance principle typical of CC networks. On the other hand, the correlation of these volumes do not correspond to an analogous correlation between the number of outgoing and incoming transactions. More specifically, the incoming transactions are very lowly correlated with the other measures. A medium-high intensity of correlation ($0.68$ and $0.59$) is observed between the outgoing transactions and volumes in $2023$. 

\begin{table}[ht]
    \centering
\resizebox{0.25\textwidth}{!}{    \begin{tabular}{rrrrrr} \hline
Year&&$\theta^{\text{out}}$&$\theta^{\text{in}}$&$v^{\text{out}}$&$v^{\text{in}}$\\\hline
\multirow{4}{*}{2022}&$\theta^{\text{out}}$&1&	0.41&	0.49&	0.50\\
&$\theta^{\text{in}}$&0.41&	1&	0.19&	0.17\\
&$v^{\text{out}}$&0.49&	0.19&	1&	0.93\\
&$v^{\text{in}}$&0.50&	0.17&	0.93&	1\\ \hline \hline 
Year&&$\theta^{\text{out}}$&$\theta^{\text{in}}$&$v^{\text{out}}$&$v^{\text{in}}$\\\hline
\multirow{4}{*}{2023}&$\theta^{\text{out}}$&1&0.23&0.68&0.59\\
&$\theta^{\text{in}}$&0.23&1&0.23&0.26\\
&$v^{\text{out}}$&0.68&0.23&1&0.93\\
&$v^{\text{in}}$&0.59&0.26&0.93&1\\\hline \hline 
Year&&$\theta^{\text{out}}$&$\theta^{\text{in}}$&$v^{\text{out}}$&$v^{\text{in}}$\\\hline
\multirow{4}{*}{2024}&$\theta^{\text{out}}$&1&	0.29&	0.51&	0.55\\
&$\theta^{\text{in}}$&0.29&	1&	0.24&	0.27\\
&$v^{\text{out}}$&0.51&	0.24&	1&	0.89\\
&$v^{\text{in}}$&0.55&	0.27&	0.89&	1\\\hline
         \end{tabular}}
    \caption{Correlations between the vectors of transactions and volumes: $\theta^{\text{out}}=\text{col}(\{\theta_i^{\text{out}}\}_{i \in \mathcal{N}})$, $\theta^{\text{in}}=\text{col}(\{\theta_i^{\text{in}}\}_{i \in \mathcal{N}})$, $v^{\text{out}}=\text{col}(\{v_i^{\text{out}}\}_{i \in \mathcal{N}})$, $v^{\text{in}}=\text{col}(\{v_i^{\text{in}}\}_{i \in \mathcal{N}})$ (2022, 2023, 2024).}
    \label{tab:correlazione}
\end{table}

\subsection{Reciprocity and cycles}

Reciprocity and cycles are graph measurements which can be useful to indicate users' behaviors which represent central features for the circulation of a CC. 

\begin{figure}[!h]
\centering
\includegraphics[width=0.8\linewidth]{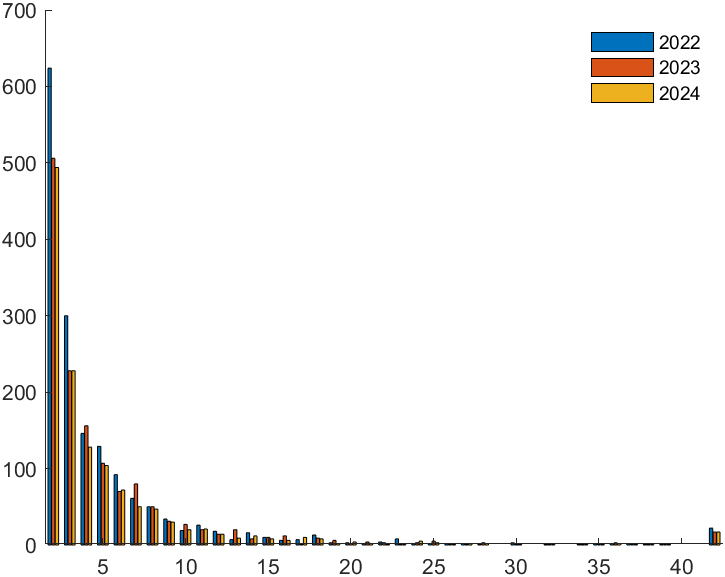}
\caption{Distribution of the reciprocity $r_i$, $i \in \mathcal{N}$ (for better graphic readability, nodes that have a single reciprocated edge are omitted, whose values can be observed in the Table~\ref{tab:reciprocity} and nodes with a reciprocity higher than 40 have been merged into
the last group of bars).}
\label{fig:recipDist}
\end{figure}

In terms of graph edges, the \textit{reciprocity} $r_i \in \{0, \dots,N\}$ of the $i$-th node, $i \in \mathcal{N}$, is defined as
\begin{equation}
 r_i = \sum_{j \in \mathcal{N}} \delta_{ij} \delta_{ji},   
\end{equation}
\noindent thus providing the number of nodes that share with the $i$-th node transactions in both directions (not necessarily the same number). Figure~\ref{fig:recipDist} shows the distribution of reciprocity over the network. The maximum value of $r_i$ in 2022 is $564$ (out of the horizontal axis limit in Figure~\ref{fig:recipDist}), while in the other two years it is almost half, specifically $283$ in 2023 and $285$ in 2024. 

Table~\ref{tab:typeReciprocity} shows the number of nodes with reciprocal bonds by different type of users.
Type B users are those who activate the greatest number of reciprocal ties, mainly with users of the same type. As for C users, reciprocal ties are recorded almost exclusively with B users. In the year 2022, some reciprocal ties are also recorded between P users.

 \begin{table}[!h]
\centering
\resizebox{0.30\textwidth}{!}{
\begin{tabular}{rrrrrr}
\hline
Year&Type&  B&  C&  E&  P\\ \hline 
\multirow{4}{*}{2022}& B&7,022&3,323&523&232\\
& C&3,323&0&0&0\\
& E&523&0&2&7\\
& P&232&0&7&2\\ \hline \hline
Year&Type&  B&  C&  E&  P\\ \hline 
\multirow{4}{*}{2023}& B&6,662&1,759&585&238\\
& C&1,759&0&0&5\\
& E&585&0&0&6\\
& P&238&5&6&0\\ \hline \hline
Year&Type&  B&  C&  E&  P\\ \hline 
\multirow{4}{*}{2024}& B&5,836&1,667&651&228\\
& C&1,667&0&0&4\\
& E&651&0&0&5\\
& P&228&4&5&0\\ \hline 
\end{tabular}
}
\caption{Reciprocal bonds distinguished by user type (2022, 2023, 2024).}
\label{tab:typeReciprocity}
\end{table}

The total number of transactions between nodes with a given value $\rho \in \{0, \dots,N\}$ of reciprocity is defined as
\begin{equation} \label{eq:tro}
            T_\rho = \sum_{i \in \mathcal{N}} \eta_{i\rho}  \sum_{j \in \mathcal{N}} \delta_{ij} \delta_{ji}(e_{ij} + e_{ji}),
        \end{equation}
where $\eta_{i\rho} \in \{0,1\}$ is given by
\begin{equation}
\eta_{i\rho} = 1 \iff \{r_i=\rho\}.  
\end{equation}
\noindent Analogously, the total volume of transactions between  nodes with a given value $\rho$ of reciprocity is defined by 
\begin{equation} \label{eq:vro}
    v_\rho = \sum_{i \in \mathcal{N}} \eta_{i\rho} \sum_{j \in \mathcal{N}} \delta_{ij} \delta_{ji} (w_{ij} + w_{ji}).
\end{equation}

\noindent Table~\ref{tab:reciprocity} shows the distributions corresponding to \eqref{eq:tro} and \eqref{eq:vro} and the number of nodes $N_\rho=\sum_{i \in \mathcal{N}} \eta_{i\rho}$ involved. It should be noticed that the values of $N_\rho$, $T_\rho$, and $v_\rho$ are typically  decreasing with $\rho$, even though this is not always the case. Moreover, the cumulative values for $\rho \ge 15$, although not very significant in terms of number of nodes, are far from negligible in terms of number of transactions and corresponding volumes. 
%


\begin{table}[!th]
\centering
\resizebox{0.5\textwidth}{!}{
\begin{tabular}{rrrrrrrrrr}\\
\hline
$\rho$ & $N_{\rho}^{(2022)}$ & $N_{\rho}^{(2023)}$ & $N_{\rho}^{(2024)}$ & $T_{\rho}^{(2022)}$ & $T_{\rho}^{(2023)}$ & $T_{\rho}^{(2024)}$ & $v_{\rho}^{(2022)}$ & $v_{\rho}^{(2023)}$ &$v_{\rho}^{(2024)}$ \\\hline
1 & 4,362 & 3,075 & 2,879 & 87k & 42k & 48k & 3,074\,k\euro & 3,332\,k\euro &3,293\,k\euro \\
2 & 624 & 506 & 494 & 9k & 6k & 6k & 1,950\,k\euro & 1,982\,k\euro &1,900\,k\euro\\
3 & 300 & 228 & 228 & 4k & 3k & 3k & 1,474\,k\euro & 917\,k\euro &915\,k\euro \\
4 & 146 & 156 & 128 & 2k & 2k & 3k & 909\,k\euro & 1,325\,k\euro &609\,k\euro \\
5 & 129 & 107 & 104 & 3k & 2k & 3k & 1,138\,k\euro & 665\,k\euro &630\,k\euro\\
6 & 92 & 70 & 72 & 3k & 2k & 2k & 728\,k\euro & 712\,k\euro &998\,k\euro \\
7 & 61 & 80 & 50 & 2k & 3k & 2k & 558\,k\euro & 775\,k\euro &533\,k\euro \\
8 & 50 & 50 & 47 & 2k & 3k & 1k & 497\,k\euro & 849\,k\euro &1,229\,k\euro\\
9 & 34 & 31 & 30 & 1k & 1k & 2k & 374\,k\euro & 223\,k\euro &306\,k\euro \\
10 & 19 & 27 & 20 & 1k & 2k & 1k & 123\,k\euro & 311\,k\euro &167\,k\euro \\
11 & 26 & 20 & 21 & 2k & 1k & 1k & 259\,k\euro & 245\,k\euro &160\,k\euro \\
12 & 18 & 14 & 14 & 1k & 900 & 600 & 399\,k\euro & 220\,k\euro &129\,k\euro \\
13 & 7 & 20 & 9 & 500 & 2k & 1k & 52\,k\euro & 149\,k\euro &118\,k\euro \\
14 & 16 & 8 & 12 & 1k & 500 & 2k & 271\,k\euro & 234\,k\euro &187\,k\euro \\
$\ge 15$ & 100 & 88 & 79 & 47k & 42k & 43k & 2,583\,k\euro & 3,000\,k\euro&2,664\,k\euro\\\hline
    \end{tabular}
}
\caption{Different values of reciprocity $\rho$ and corresponding: number of nodes $N_\rho$, number of transactions $T_\rho$, and volume $v_\rho$ (2022, 2023, 2024).}
\label{tab:reciprocity}
\end{table}

Cycles represent a circular relationship among users of the circuit, which is a desirable behavior in the circulation of a CC. 
Table~\ref{tab:cycles} indicates some characteristics of the cycles of length $\ell \in \{2,3,4,5\}$.  
Both the number of nodes participating in cycles of length $\ell$ ($N_{n\ell}$)  and the total number of cycles of length $\ell$ ($N_{c\ell}$) are always greater in 2022 than in 2023 and 2024. This is also true for the number of nodes participating in a single cycle ($N_{n\ell_1}$) and for the maximum number of cycles in which a single node can participate ($N_{c,\max}$), but exclusively for $\ell=2$ and $\ell=3$ in the first case, and for $\ell=3$ and $\ell=5$ in the second case.
Belonging to a cycle would seem to be a less desirable condition for circuit users. This behavior could reflect their preference for more flexible, dynamic, and reciprocal relationships, thus avoiding being tied to rigid or repetitive patterns.

 \begin{table}[ht]
     \centering
     \resizebox{0.35\textwidth}{!}{
     \begin{tabular}{rrrrrr} \hline 
     Year &$\ell$ & $N_{n\ell}$ & $N_{c\ell}$ & $N_{n\ell_1}$& $N_{c,\max}$ \\ \hline
     \multirow{4}{*}{2022}& 2&5,984&7,598&4,362&564\\
     &3&4,282&16,623&1,370&2,548\\
     &4&8,696&604,291&470&320,130\\
     &5&9,271&4,425,354&112&2,207,172\\\hline\hline
      Year &$\ell$ & $N_{n\ell}$ & $N_{c\ell}$ & $N_{n\ell_1}$& $N_{c,\max}$ \\ \hline
\multirow{4}{*}{2023}&2&4,480&5,924&3,075&283\\
&3&4,033&16,317&1,283&2,875\\
&4&7,242&352,559&574&104,636 \\
&5&7,847&4,505,744&126&2,595,034\\ 
\hline\hline
 Year &$\ell$ & $N_{n\ell}$ & $N_{c\ell}$ & $N_{n\ell_1}$& $N_{c,\max}$ \\ \hline
 \multirow{4}{*}{2024}& 2&4,187&5,473&2,879&285\\
     &3&3,713&14,357&1,170&2,321\\
     &4&6,501&317,160&423&81,682\\
     &5&7,055&3,405,815&166&1,886,248\\\hline
 \end{tabular}}
\caption{Some characteristics of the cycles of length from $2$ to $5$ (2022, 2023, 2024): $\ell$ is length of the cycle, $N_{n\ell}$ is the number of nodes participating to  cycles of length $\ell$, $N_{c\ell}$ is the number of cycles of length $\ell$, $N_{n\ell_1}$ is the number of nodes that participate to a single cycle of length $\ell$, $N_{c,\max}$ is the maximum number of cycles of length $\ell$ in which a single node is involved.}
\label{tab:cycles}
\end{table}

\subsection{Local clustering and transitivity}

Clustering behavior is a further key aspect for monetary networks. In this section, local and global clustering measurements are analyzed in the perspective of the Sardex network. In addition, other clustering based on geolocalization of Sardex users and their business sector are reported in the Appendix.

\begin{figure}
    \centering
    \includegraphics[width=0.8\linewidth]{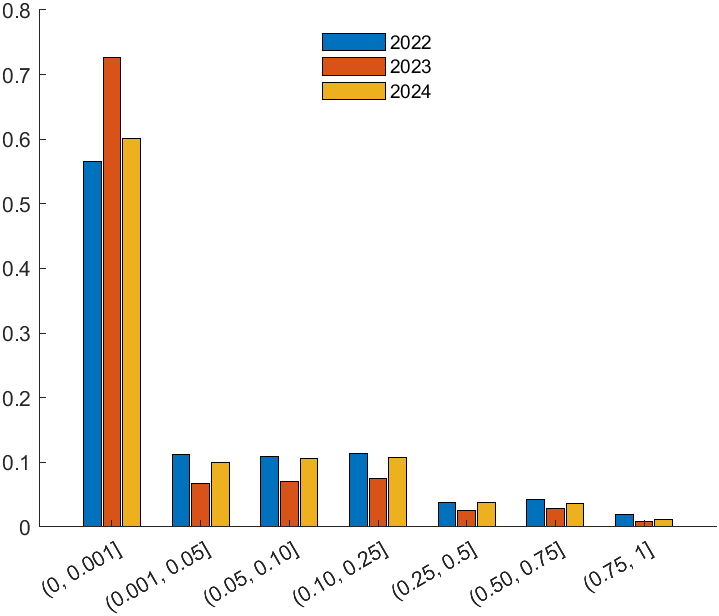}
    \caption{Distribution of the local clustering coefficient of users.}
    \label{fig:istCC}
\end{figure}

The local clustering coefficient is the density of the ego-network of each node and can be measured as the ratio between the number of edges among the neighbors of the node and its maximum possible value. The distribution of the local clustering coefficient is shown in Figure~\ref{fig:istCC}. The average tendency of nodes to form clusters, quantified as the average of local cluster coefficients, is equal to $8.1\%$, $4.9\%$, and $6.9\%$ respectively in $2022$, $2023$, and $2024$. Most nodes of the network 
($56.5\%$ in $2022$, $72.6\%$ in $2023$, $60.0\%$ in $2024$)  have a zero value of the clustering coefficient, few ($6.2\%$ in $2022$, $3.7\%$ in $2023$, $4.9\%$ in $2024$) 
have an index larger than $0.5$, and very few ($1.8\%$ in $2022$, $0.7\%$ in $2023$, $1.1\%$ in $2024$)
show a clustering coefficient equal to one. 

\begin{figure}[ht]
    \centering
    \includegraphics[width=0.8\linewidth]{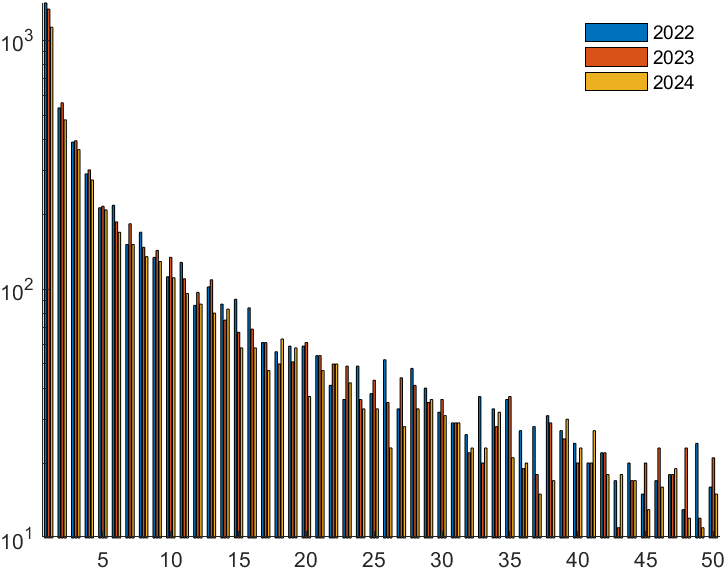}
    \caption{Nodes (vertical axis, scale is set to logarithmic scale) participating to a certain number $n_{\Delta}$ of triangles  (horizontal axis).}
    \label{fig:barTriangoli}
\end{figure}

\begin{figure}[ht]
    \centering
    \includegraphics[width=0.8\linewidth]{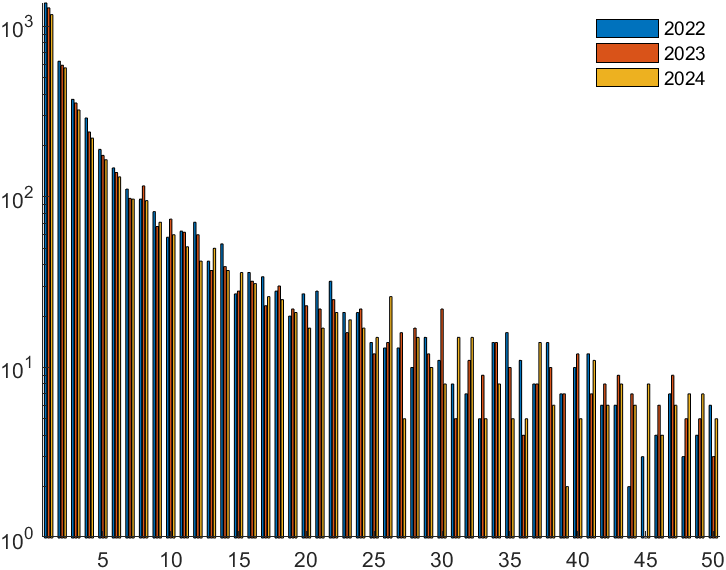}
    \caption{Nodes (vertical axis, scale is set to logarithmic scale) participating to a certain number $n_\Delta^{\text{sc}}$ of strongly connected triplets.}
    \label{fig:barTriangoliSC}
\end{figure}

In order to consider the clustering over triplets of the graph, one can consider the number $n_\Delta$ of closed triplets (called triangles) in the undirected graph, the number $n_{\Delta}^{\text{sc}}$ of strongly connected triplets in the directed graph (obviously each strongly connected triplet is also a triangle but not vice versa), and the number $n_\Lambda$ of connected (open and closed) triplets in the undirected graph. 
The distributions of $n_{\Delta}$  and $n_{\Delta}^{\text{sc}}$ are shown in Figure~\ref{fig:barTriangoli} and Figure~\ref{fig:barTriangoliSC}, respectively. 
The values of $n_\Delta$ are very close to  $2n_{\Delta}^{\text{sc}}$ which means that most of the connected triplets are strongly connected.    

Most of the nodes do not participate in any triads: $56.5\%$ in $2022$, $72.6\%$ in $2023$, and $60.3\%$ in $2024$ of the nodes do not participate to triangles, and $70.8\%$ in $2022$, $82.2\%$ in $2023$, and $72.9\%$ in $2024$ of the nodes do not participate to strongly connected triplets. 
The nodes involved in closed triplets, possibly strongly connected, play an important role in the monetary movement inside the network. 
The user who participates in the greatest number of triangles is one of the users of type P in all three years analyzed, however the number of triangles in which it participates is less than 1\% of the possible triangles that can be activated. The same result is observed if we consider the strongly connected triplets.

\section{Condensed graph for  community detection}
\label{sec:cond}
Understanding the structural backbone of a transactional network is essential for identifying the mechanisms that sustain or hinder money circulation. In the context of CCs like Sardex, where credit is expected to recirculate through user interactions, analyzing the SCCs provides insight into the resilience and cohesion of the system. This section introduces a condensed graph representation of the Sardex network, where each node corresponds to an SCC, and the resulting directed acyclic graph (DAG) reveals mesoscopic connectivity patterns. We begin by formalizing the relevant graph-theoretic notation, then apply this framework to examine the evolution, structural distribution, and economic relevance of the core structural components over a three-year period.

\subsection{Graph condensation}

The condensation of a digraph $\mathcal{G}$ is a DAG $\mathcal{G}^C = (\mathcal{N}^C, \mathcal{E}^C)$, where each node in $\mathcal{N}^C$ represents an SCC of $\mathcal{G}$, and edges represent connections between these SCCs. Specifically, if there exists at least one edge from a node in the $i$-th SCC to a node in the $j$-th SCC in the original graph, then there exists an edge from node $i$ to node $j$ in the condensation graph. 

For synthetic graphs, when the set $\mathcal{N}$ is infinite, we can define the giant weakly connected component (GWCC) as the only (if it exists) weakly connected component with infinite dimension. Analogously, we can define the giant strongly connected component (GSCC) as the only SCC  (if it exists) with infinite dimension. 
Differently from the case of synthetic networks, for real-world networks, the threshold size for the largest SCC to be considered a GSCC has not been rigorously defined: it is denoted as the largest SCC that contains a significant fraction of the entire graph's vertices. 

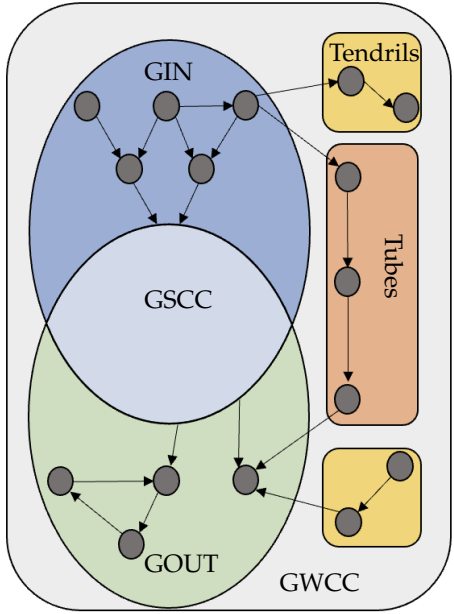
\begin{figure}[ht]
    \centering
   \begin{overpic}[
   width=0.7\columnwidth]{
   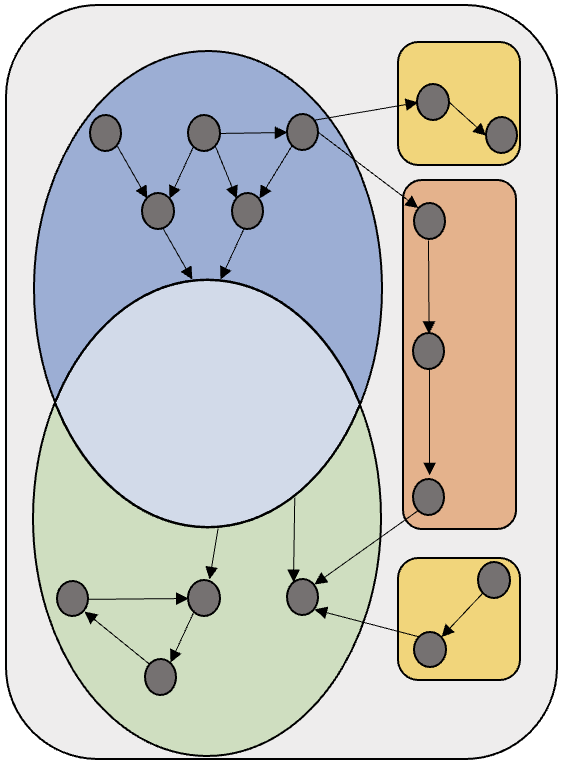}
        \put(23, 87){GIN}
        \put(23, 50){GSCC}
        \put(23, 7){GOUT}
        \put(53, 90){Tendrils}
        \put(62, 62){\rotatebox{270}{Tubes}}
        \put(45, 4){GWCC}
    \end{overpic}
    \caption{A GWCC representation with its components.
    }
    \label{fig:condensation}
\end{figure}

When both the GWCC and the GSCC exist, then, of course, the former encompasses the latter. In this case, within the GWCC, we can distinguish the giant in-component (GIN) and the giant out-component (GOUT) that, respectively, are the upstream and downstream of the GSCC including it and whose intersection is the GSCC itself, see Figure \ref{fig:condensation} for a visual representation. Then, the graph is completed by the tubes, namely the nodes not in the GSCC that are encompassed in directed paths originating in the GIN and ending in the GOUT, and finally the tendrils that are the remaining SCCs in the GWCC \cite{dorogovtsev2001giant,ancona2024percolation} that are neither in the GIN or GOUT.
\subsection{Sardex condensation}
%




To understand the peculiarities of the Sardex network at a mesoscopic scale, we analyzed the evolution of its condensation over the years 2022, 2023, and 2024, by applying the algorithm \cite{hopcroft1973algorithm} to the its GWCC (see Table~\ref{tab:with_management}). 
\begin{table}[ht]
    \centering
    \resizebox{0.95\columnwidth}{!}{
    \begin{tabular}{lcccc}
        \hline
        Year & GWCC & GSCC & GIN\textbackslash GSCC & GOUT\textbackslash GSCC \\
        \hline
        2022 & 14,649 &64.4\% & 23.0\% & 12.3\% \\
        2023 & 22,657 & 35.7\% & 11.8\% & 52.3\% \\
        2024 & 13,704 & 52.5\% & 11.4\% & 35.8\% \\
        \hline
    \end{tabular}
    }
    \caption{Network statistics including provider nodes (2022, 2023, 2024).}
    \label{tab:with_management}
\end{table}
In all three years the nodes that belong to the GWCC and do not belong to either the GIN or GOUT, i.e., tendrils and tubes,  are negligible ($0.2\%$ in 2023 and $0.3\%$ in 2022 and 2024), indicating that most nodes were either actively engaged in transactions or linked to core network components.

In 2023 and 2024, the Sardex network exhibits a structure that is unbalanced in the sense that the size of the GOUT substantially outweighs that of the GIN. We posit that this is due to the activity of the provider nodes, whose transactions distort the network structure by creating a high GOUT proportion, and ensuring that many users receive credits but do not contribute to cyclical transaction flows.


The network expansion in 2023 is relevant, which could suggests an increase in participation, likely driven by heightened adoption of Sardex transactions. However, the decline in the size of the GSCC suggests that while many new nodes were integrated into the network, this has not resulted in reciprocal transactional connectivity within the core structure, as testified by the size of GOUT\textbackslash GSCC.


\subsection{Transaction and volume weighted condensations}

Analyzing the Sardex network based on the number of transactions provides insights into how economic activity is distributed across different network components. Unlike node-based analyses, this method highlights the frequency of credit circulation rather than the number of participants. Table~\ref{tab:weighted_network} presents the transaction-weighted proportions of key network components from 2022 to 2024.

\begin{table}[ht]
    \centering
    \resizebox{0.8\columnwidth}{!}{
    \begin{tabular}{lccc}
        \hline
        Year & GSCC & GIN\textbackslash GSCC & GOUT\textbackslash GSCC \\
           \hline
        2022 & 94.8\% & 2.7\% & 2.5\% \\
        2023 & 89.6\% & 2.1\% & 8.3\% \\
        2024 & 92.8\% & 1.7\% & 5.5\% \\
        \hline
    \end{tabular}}
    \caption{Transaction-weighted proportions of key network components (2022, 2023, 2024).}
    \label{tab:weighted_network}
\end{table}

The GSCC consistently processes the majority of transactions, even as its dominance fluctuates. In 2022, $94.8\%$ of transactions occurred within the GSCC, indicating that most economic activity remained within the core. This proportion dropped in 2023 to $89.6\%$, suggesting a temporary increase in transactions involving the GOUT. By 2024, the GSCC’s transactional share recovered to $92.8\%$, reflecting renewed stability in credit circulation within the core. The number of transactions within the GIN and GOUT excluding the GSCC constitutes a small amount, even though there is a positive trend in GOUT\textbackslash GSCC from 2022 to 2024, compared to a negative trend in GIN\textbackslash GSCC. Tendrils and tubes contribute minimally to transaction frequency. In all years, tendrils account for less than $0.03\%$ of transactions, confirming their marginal role in economic circulation. Remarkably, in 2023, the GSCC handled a lower proportion of transactions compared to 2022 and 2024. This aligns with previous findings that credits in 2023 flowed towards the GOUT without significant reinvestment.

Beyond the number of transactions, analyzing the network based on transaction volumes reveals how economic value circulates. Table~\ref{tab:volume_weighted} presents the volume-weighted proportions of the Sardex network.

\begin{table}[ht]
    \centering
    \resizebox{0.8\columnwidth}{!}{
    \begin{tabular}{lccc}
       \hline
        Years & GSCC & GIN\textbackslash GSCC & GOUT\textbackslash GSCC \\
        \hline
        2022 & 97.6\% & 0.8\% & 1.7\%  \\
        2023 & 97.8\% & 0.6\% & 2.2\%  \\
        2024 & 96.6\% & 0.7\% & 2.7\% \\
        \hline
    \end{tabular}
    }
    \caption{Volume-weighted proportions of key network components (2022, 2023, 2024).}
    \label{tab:volume_weighted}
\end{table}

The GSCC remains the dominant structure in terms of transaction volume. Across all the three years, about $97\%$ of the total value transferred occurred within the GSCC. The GIN and GOUT maintain marginal levels of economic activity, with a slight increase of the volume exchanged within the GOUT\textbackslash GSCC. 

Both weighting approaches confirm that the GSCC remains the core of economic transactions within the Sardex network. 
Over time, the relevance of the GSCC has slightly decreased both in terms of number and volume of transactions, suggesting a trend toward more decentralized economic activity. These findings reinforce the crucial role of the structural graph-theoretical techniques to investigate the health of the CC network and the need for further research into reinvestment behaviors and policy strategies to sustain credit circulation within the core network.

\subsection{Community structures: a close-up view}

\begin{figure}
\centering
\includegraphics[width=0.15\textwidth]{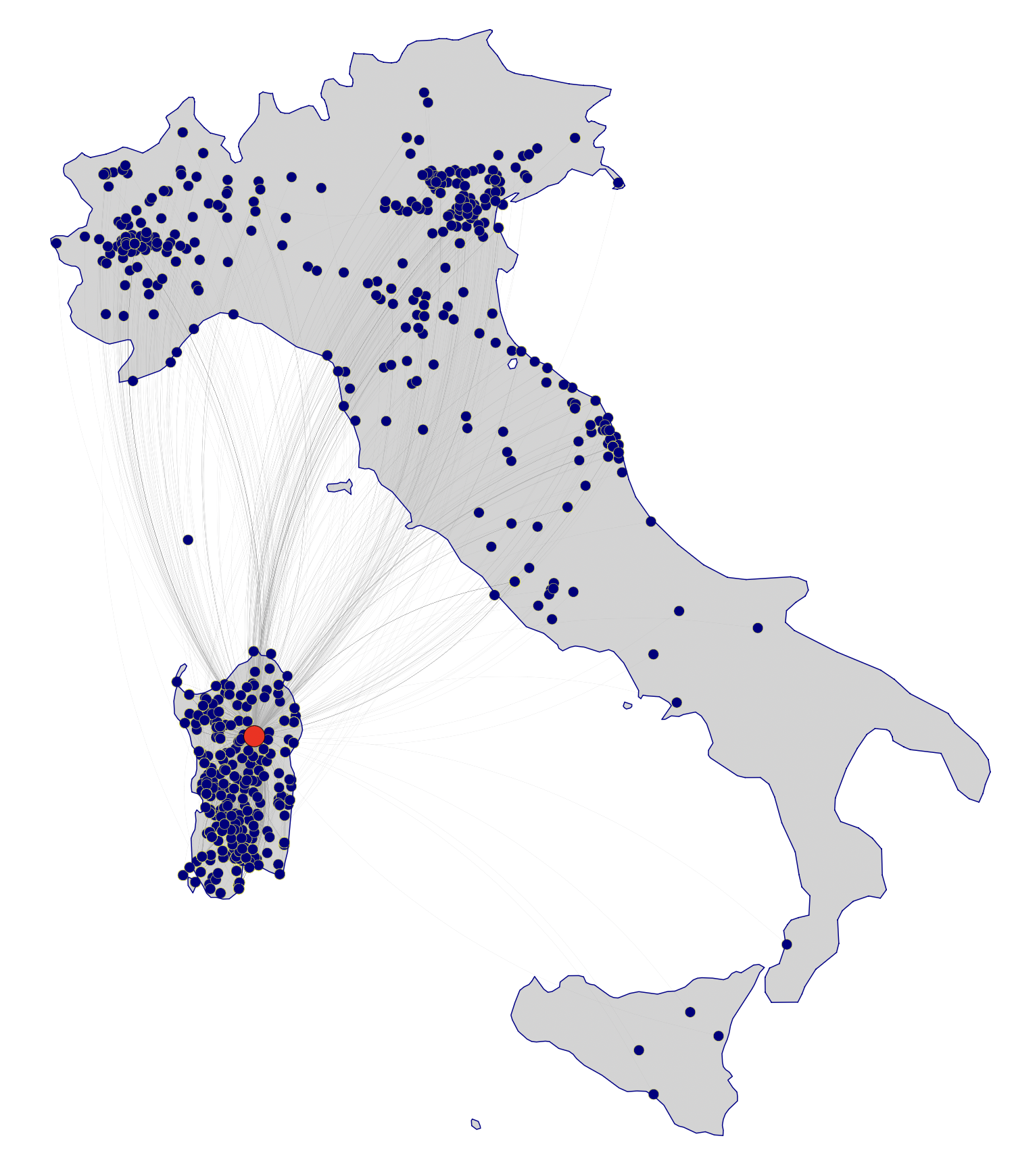}
\includegraphics[width=0.15\textwidth]{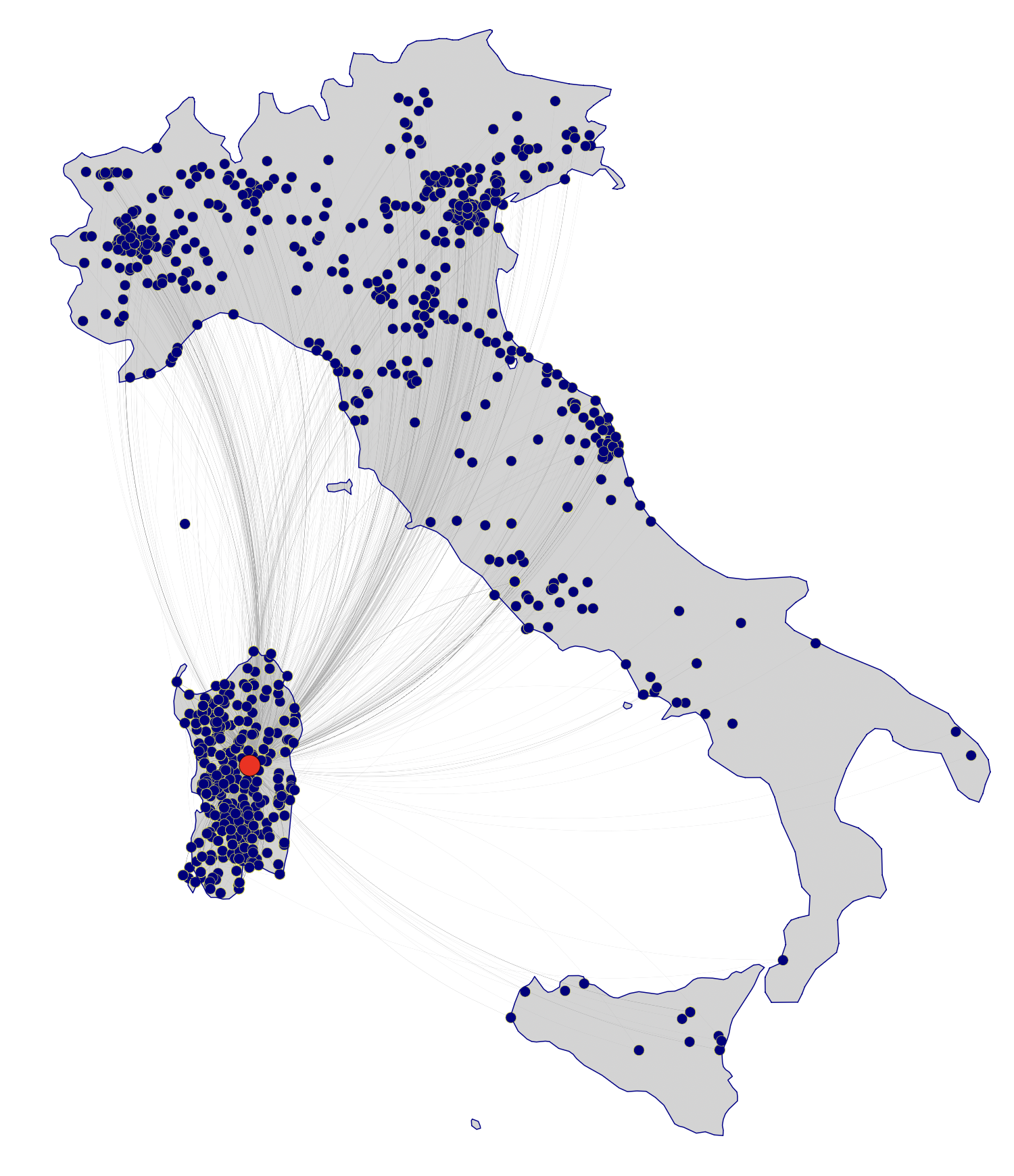}
\includegraphics[width=0.15\textwidth]{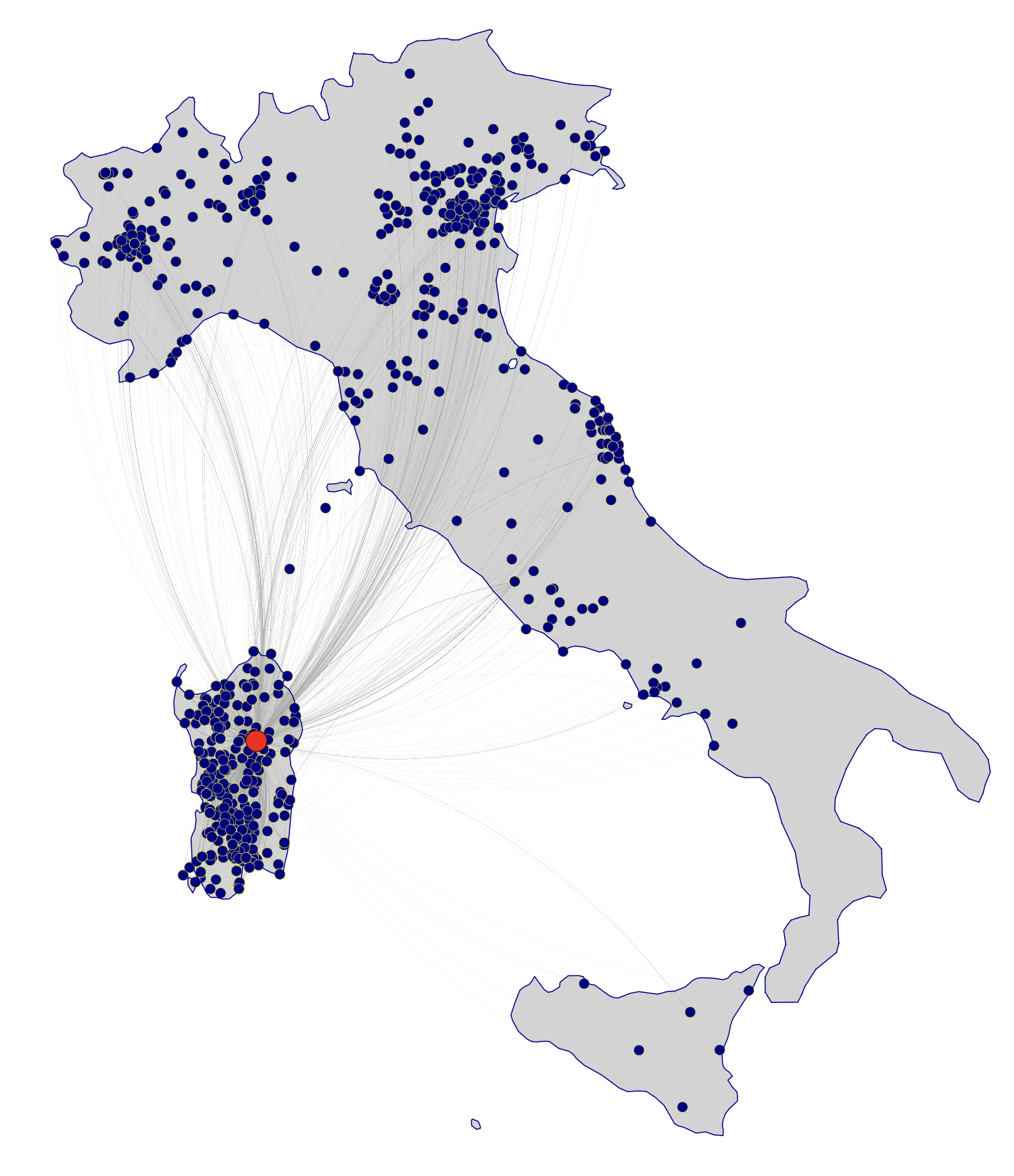}
\caption{Georeferenced representation of the spatial distribution of the SCCs within the condensed graph (GSCC is the red point located in Sardinia), where the coordinates of each SCC are computed as the average of the coordinates of the nodes that constitute it, for years 2022 (left), 2023 (center), and 2024 (right).} 
\label{fig:georef_cond_map}
\end{figure}

A geographical assessment of the condensed graph is shown in Figure \ref{fig:georef_cond_map}. The spatial placement of nodes is  determined by averaging the coordinates of their original network counterparts. The representation confirms that the GSCC remains predominantly located in northern Sardinia, reinforcing previous findings. The presence of few nodes located in the sea (one in 2022 and 2023, two in 2024) demonstrates the lack of strong interconnections between the island and external regions.

The Sardex network includes at most four provider (P), see Table~\ref{tab:typememb}. Let us consider how the P  users influence the network transactional structure. In 2023, two of the P users (Serramanna and Padova) belong to the GSCC, actively engaging in transactions, while the other two (Aosta and Albiate) are part of the GOUT. These nodes generate a significant number of transactions, primarily flowing outward to nodes in the GOUT, which biases the network’s structure by increasing the number of participants who receive credits without reinvesting them. To assess their impact, we analyze the GSCC, GIN, and GOUT across 2022, 2023, and 2024, after filtering out transactions that originate from or go towards P users.

Removing the providers   transactions (Table~\ref{tab:without_management}) reveals several important trends (compare with Table~\ref{tab:with_management}): 
\begin{itemize} 
\item By excluding the P users, the GSCC remains the dominant structure in organic transactions across all years.
\item Filtering out provider transactions, the GIN and GOUT become more proportionate. In 2023, the GOUT was heavily inflated ($88.0\%$) due to provider  nodes, but after filtering, it dropped significantly to $76.6\%$. A similar trend occurred in 2024, where the GOUT dropped from $88.3\%$ to $82.0\%$, demonstrating that provider transactions distort the apparent distribution of credit flow.
\item The filtered network GWCC in 2024 was the smallest (10,323 nodes) compared to 11,836 (2023) and 14,072 (2022). This decline in organic transactions indicates that fewer users were engaging in the system without external credit inflows, potentially signaling a weakening of reinvestment cycles. 
\end{itemize}

\begin{table}[ht]
    \centering
    \resizebox{\columnwidth}{!}{
    \begin{tabular}{lcccc}
        \hline
        Years & GWCC & GSCC & GIN\textbackslash GSCC & GOUT\textbackslash GSCC \\
        \hline
        2022 & 14,072 & 64.4\% & 24.7\% & 10.1\% \\
        2023 & 11,836 & 64.2\% & 22.8\% & 12.4\% \\
        2024 & 10,323 & 65.8\% & 16.9\% & 16.2\% \\
        \hline
    \end{tabular}
    }
    \caption{Network statistics excluding P users (2022, 2023, 2024).}
    \label{tab:without_management}
\end{table}

\section{Comparison with null model}
\label{sec:nm}
To assess the structural distinctiveness of the Sardex network and the behavioral interpretation of its observed features, we perform a series of comparisons of the binary Sardex graphs against a randomized null model. This model preserves in-degree and out-degree distributions, while randomizing edge configurations \cite{barabasi2003scale,newman2001random}. The mean values of the variables over $50$ runs are taken to remove peculiar patterns, if present. By comparing the empirical network’s condensation structure and behavioral patterns with those expected under degree-preserving randomization, we aim to identify which structural features arise from non-random organizing principles—such as imitation—that cannot be captured by the degree distribution alone. 

\subsection{Imitation and out-degree preferences}
The comparison of the Sardex network with the null model confirms the statistical significance of the results (Kolmogorov-Smirnov test with $p<0.01$) and highlights that links between Sardex users are not random, but rather result from specific structural and behavioral patterns. In particular,  in the Sardex circuit, users tend to connect with others who are more active than them as demonstrated by the following analysis. 

\begin{figure}
    \centering
    \includegraphics[width=\linewidth]{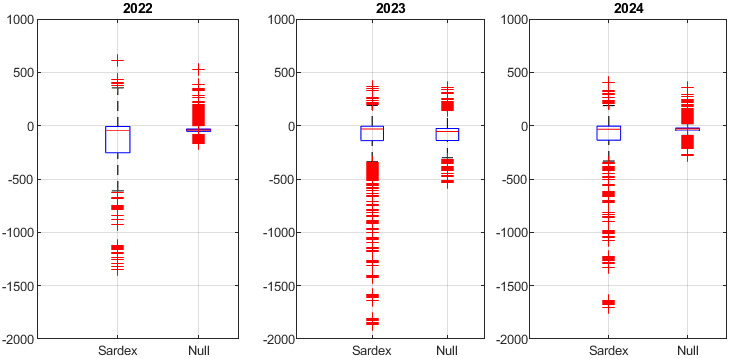}
 \caption{Boxplots of the out-degree gaps \eqref{eq:gi} for the Sardex data and the null model in the years 2022 (left), 2023 (middle), and 2024 (right).}
    \label{fig:out_degree_gap}
\end{figure}

Let's consider the difference between the number of (outgoing, omitted below) neighbors of the $i$-th user, i.e., $|\mathcal{N}_i|=\sum_{j \in \mathcal{N}} \delta_{ij}$, with the average number of neighbors of its neighbors:
\begin{equation}
  \Delta_i^{\text{out}} = |\mathcal{N}_i| - \frac{1}{|\mathcal{N}_i|}\sum_{j\in \mathcal{N}_i}|\mathcal{N}_j|, \label{eq:gi}
\end{equation}
for all $i \in \mathcal{N}$. The users such that $\mathcal{N}_i=\emptyset$, , $i \in \mathcal{N}$, are excluded from the computation of \eqref{eq:gi} as well as for the other indices. The analysis of the boxplots of \eqref{eq:gi} shown in Figure~\ref{fig:out_degree_gap} reveals a negative asymmetry which is due to the non-negligible percentage of nodes with high number of neighbors. On the other hand, some peculiarities of the Sardex circuit can be highlighted. In particular, especially in the years 2022 and 2024, the larger boxes demonstrate that Sardex users tend to connect with users who have more neighbors than theirs, more often than would be expected under the null model. 


    \begin{figure}
    \centering
    \includegraphics[width=\linewidth]{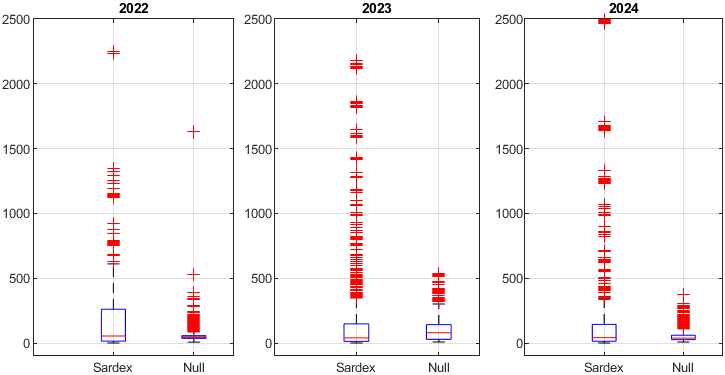}
 \caption{Boxplots of $\Delta_i^{\text{av}}$, $i \in \mathcal{N}$, given by    \eqref{eq:mean} for the Sardex data and the null model in the years 2022 (left), 2023 (middle), and 2024 (right).}
    \label{fig:out_degree_deltamean}
\end{figure}

A similar behavior is highlighted by considering the average of the difference between the number of neighbors of the $i$-th user and those of its neighbors, which is defined as
\begin{equation}
    \Delta^{\text{av}}_i=\frac{1}{|\mathcal{N}_i|}\sum_{j \in \mathcal{N}_i} \left| \,|\mathcal{N}_i| - |\mathcal{N}_j| \, \right|, \label{eq:mean}
\end{equation} 
for all $i \in \mathcal{N}$ except those for which $\mathcal{N}_i=\emptyset$. The boxplot of \eqref{eq:mean} for the Sardex data and the null model are shown in Figure~\ref{fig:out_degree_deltamean}: in the null model, the mean deviation is more contained and concentrated on low values with respect to the Sardex network. 

\begin{figure}
    \centering
\includegraphics[width=\linewidth]{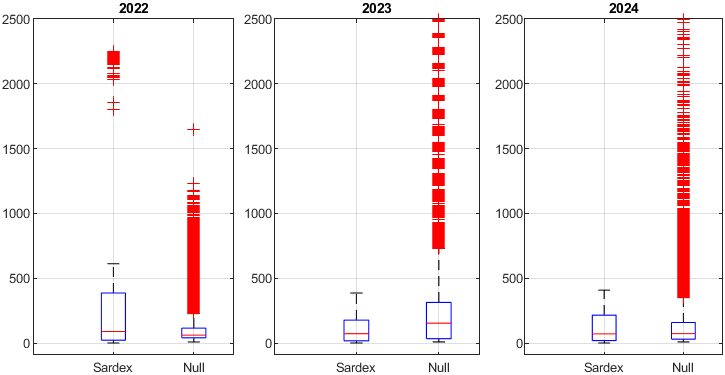}
    \caption{Boxplots of $\Delta_i^{\text{max}}$, $i \in \mathcal{N}$, given by \eqref{eq:max} for the Sardex data and the null model in the years 2022 (left), 2023 (middle), and 2024 (right). For the years 2023 and 2024, several upper outliers appearing in the Sardex data have not been represented for the sake of readability.}
    \label{fig:deltaMax}
\end{figure}

The interpretation proposed above is confirmed by the analysis of the maximum difference of neighbors between the $i$-th user and its neighbors, which can be defined as
\begin{equation}
    \Delta^{\max}_i=\max_{j \in \mathcal{N}_i}\left| \,|\mathcal{N}_i| - |\mathcal{N}_j| \, \right|. \label{eq:max}
\end{equation}
The boxplots reported in Figure~\ref{fig:deltaMax} show that, especially for 2022 and 2024, the real  distribution tends to be wider than the randomized one, indicating that in the Sardex network there are nodes with higher local variations than in the null model. The structure of the real network seems to favor connections between nodes with significant differences in their number of neighbors.

\begin{figure}
    \centering
\includegraphics[width=\linewidth]{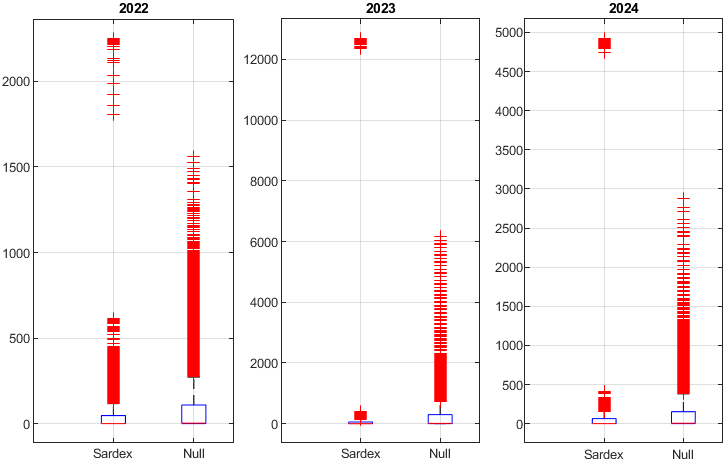}
 \caption{Boxplots of $\Delta_i^{\text{conf}}$, $i \in \mathcal{N}$, given by \eqref{eq:ci} for the Sardex data  and the null model in the years 2022 (left), 2023 (middle), and 2024 (right).}
    \label{fig:intConf}
\end{figure}

The analysis of the confidence intervals shows another behavior: Sardex users' neighbors are more similar to each other than in randomized connections. Let us define the confidence interval of the $i$-th user as 
\begin{equation}
    \Delta_i^{\text{conf}} = \max_{j \in \mathcal{N}_i} |\mathcal{N}_j|-\min_{j \in \mathcal{N}_i} |\mathcal{N}_j|, \label{eq:ci}
\end{equation}
\noindent which represents  the maximum number of neighbors minus the minimum number of neighbors among  the neighbors of the $i$-th user. Smaller values of $\Delta_i^{\text{conf}}$ suggest stronger alignments, indicating that the neighbors of the $i$-th user are more homogeneous. The boxplots of~\eqref{eq:ci} reported in Figure~\ref{fig:intConf} show that, compared to the null model, a larger portion of users have relatively small confidence intervals, suggesting some homogeneity among the degrees of proximity of Sardex users. 

In summary, Sardex users tend to interact with other users who have more neighbors than their own, highlighting a hierarchical and disassortative structure. However, the number of neighbors of the neighbors of each Sardex node are relatively homogeneous among themselves, suggesting a certain internal coherence in the local contexts of the network.

\subsection{Comparison of condensation structure}


We now turn our attention towards the mesoscopic scale, and compare the DAG condensation of the Sardex network with that of the null model. 
%
%
The results of this comparison are shown in Tables~\ref{tab:comp_null} and~\ref{tab:cond_nogest}, both when including and excluding the provider nodes and their associated transitions. In both scenarios, the observed structures show a generally strong correspondence with the null model across all components, indicating that much of the network's organization can be attributed to its degree distribution. However, the GIN, GOUT, and GSCC are slightly larger in the real network than in the corresponding null model, with the exception of the GIN of 2024 in Table~\ref{tab:cond_nogest}. The larger GSCC suggests that nodes tend to organize themselves into a highly connected core, well beyond what would be expected with degree-preserving  random configurations.  This tendency persists even when the provider nodes and their transitions are excluded, indicating that the presence of a strongly connected core reflects a structural property of the network, rather than being solely driven by the activities of the providers. 
\begin{table}[ht] 
\centering
\scalebox{0.7}{
\renewcommand{\arraystretch}{1.2}
\begin{tabular}{rrrr}
\hline
Year & Structure & Sardex Network & Null Model \\
\hline
\multirow{8}{*}{2022} &Nodes & 14,649 & 14,649 \\
&GIN & 87.4\% & 85.3\% \\
&GIN\textbackslash GSCC & 23.0\% & 23.4\% \\
&GOUT & 76.7\% & 75.0\% \\
&GOUT\textbackslash GSCC & 12.3\% & 13.1\% \\
&GSCC & 64.4\% & 62.9\% \\
&Tendrils &0.3\% & 1.1\% \\
&Tubes & 1.0\% & < 0.01\% \\
\hline\hline
Year & Structure & Sardex Network & Null Model \\
\hline
\multirow{8}{*}{2023} &Nodes & 22,657 & 22,657 \\
&GIN & 47.5\% & 42.4\% \\
&GIN\textbackslash GSCC & 11.8\% & 10.3\% \\
&GOUT & 88.0\% & 85.7\% \\
&GOUT\textbackslash GSCC & 52.3\% & 53.6\% \\
&GSCC & 35.7\% & 32.1\%\\
&Tendrils & 0.2\% & 1.7\% \\
&Tubes & 0.5\% & < 0.01\% \\
\hline \hline
Year & Structure & Sardex Network & Null Model \\
\hline
\multirow{8}{*}{2024} &Nodes & 13,704 & 13,704 \\
&GIN & 63.9\% & 59.9\% \\
&GIN\textbackslash GSCC & 11.4\% & 10.7\% \\
&GOUT & 88.3\% & 86.9\%\\
&GOUT\textbackslash GSCC & 35.8\% & 37.8\% \\
&GSCC & 52.5\% & 49.1\% \\
&Tendrils & 0.3\% & 1.2\% \\
&Tubes & 0.8\% & <0.01\% \\
\hline
\end{tabular}
}
\caption{Comparison of condensation structures of the Sardex network and the null model (2022, 2023, 2024).}\label{tab:comp_null}
\end{table}

A comparison between Tables~\ref{tab:comp_null} and~\ref{tab:cond_nogest} shows that the condensation structures for the years 2023 and 2024 differ significantly depending on whether the provider nodes are included or excluded. This difference results from the large variation in the number of nodes involved in each setting (see Tables~\ref{tab:with_management} and~\ref{tab:without_management}), which influences the resulting graph structure. The structural shifts imply that the provider nodes act as key intermediaries that balance directional flow within the network, especially between input- and output-dominated regions, namely the GIN and GOUT. Their removal leads to a redistribution of nodes across the condensation components. Nevertheless, the robustness of the GSCC after removal of the provider nodes suggests a resilient core of reciprocal trade relations that does not rely exclusively on the providers for cohesion.

\begin{table}[ht]
\centering
\scalebox{0.7}{
\renewcommand{\arraystretch}{1.2}
\begin{tabular}{rrrr}
\hline
Year & Structure & Sardex Network & Null Model \\
\hline
\multirow{8}{*}{2022} &Nodes & 14,086 & 14,086 \\
&GIN & 89.1\% & 87.2\% \\
&GIN\textbackslash GSCC & 24.7\% & 25.3\% \\
&GOUT & 74.4\% & 72.9\%\\
&GOUT\textbackslash GSCC & 10.1\% & 10.9\% \\
&GSCC & 64.4\% & 61.9\% \\
&Tendrils & 0.7\% & 1.4\% \\
&Tubes & 1.5\% & < 0.1\% \\
\hline
\hline
Year & Structure & Sardex Network & Null Model \\
\hline
\multirow{8}{*}{2023} &Nodes & 11,812 & 11,812 \\
&GIN & 86.8\% & 84.7\% \\
&GIN\textbackslash GSCC & 22.8\% & 22.4\% \\
&GOUT & 76.7\% & 75.8\% \\
&GOUT\textbackslash GSCC & 12.4\% & 13.5\% \\
&GSCC & 64.3\% & 62.3\% \\
&Tendrils & 0.7\% &1.3\% \\
&Tubes & 1.4\% & < 0.1\% \\
\hline
\hline
Year & Structure & Sardex Network & Null Model \\
\hline
\multirow{8}{*}{2024} &Nodes & 10,339 & 10,339 \\
&GIN & 82.5\% & 84.8\% \\
&GIN\textbackslash GSCC & 16.9\% & 22.4\% \\
&GOUT & 81.9\% & 75.9\% \\
&GOUT\textbackslash GSCC & 16.2\%& 13.5\% \\
&GSCC & 65.7\% & 62.4\%\\
&Tendrils &1.1\% & 1.3\% \\
&Tubes & 1.1\% & < 0.1\% \\
\hline
\end{tabular}
}
\caption{Condensation structure of the Sardex network and null model (2022, 2023, 2024) after removal of the provider nodes.}\label{tab:cond_nogest}
\end{table}



\section{Multilayer network}
\label{sec:multi}
To build the multilayer network, we first classify users into two main distinct layers: businesses and persons, based on their user type information. Regarding the business layer, we considered the provider  nodes as businesses, and regarding the people-layer we aggregated consumers and employees, for the sake of simplicity. 
Transactions between users of the same type form intra-layer edges: business-to-business (B2B) transactions within the business layer and consumer-to-consumer (C2C) transactions within the consumer layer. Transactions between different types (B2C and C2B) constitute inter-layer edges connecting the two layers. Each edge is weighted by the total transaction volume aggregated across all exchanges between the same pair of users during the year. The resulting multilayer graph is thus directed, weighted, and composed of two layers corresponding to user type, allowing for a detailed analysis of intra- and inter-layer structural dynamics.

Table~\ref{tab:multilayer_comparison} details a comparative summary of key multilayer network metrics for the years 2022, 2023, and 2024. For each year, we report average structural properties for business and consumer layers. Specifically, we measure the average number of transactions per user (total degree), average in-degree, average out-degree, and the average transaction volume for intra-layer transactions. Additionally, we report the size of the largest SCC (GSCC) and the GIN and GOUT  within each layer.

\begin{table}[ht]
\centering
\resizebox{0.8\columnwidth}{!}{
\begin{tabular}{lccc}
\hline
Business Layer & 2022 & 2023 & 2024 \\
\hline
Number of nodes      & 5,463 & 5,347 & 4,746 \\
Avg In-Degree        & 10.7 & 10.1& 9.9 \\
Avg Out-Degree       & 13.0 & 14.6 & 13.4 \\
Avg Total Degree     & 23.7 & 24.7& 23.3 \\
Avg B2B Volume       & 976 \euro & 1,044 \euro  & 1,054 \euro  \\
Sum B2B Volume       & 46,964\,k\euro & 47,739\,k\euro & 42,666\,k\euro \\
GSCC                 & 71.5\% & 68.4\% & 68.6\% \\
GIN\textbackslash GSCC     & 6.2\%  & 5.5\%  & 5.6\% \\
GOUT\textbackslash GSCC    & 21.9\% & 25.7\% & 25.5\% \\
\addlinespace
\hline\hline
Consumer Layer & 2022 & 2023 & 2024 \\
\hline
Number of nodes      & 9,185 & 17,310 & 8,958 \\
Avg In-Degree        & 2.5 & 1.9 & 2.6 \\
Avg Out-Degree       & 1.1 & 0.5 & 0.7 \\
Avg Degree           & 3.6 & 2.4 & 3.3 \\
Avg C2C Volume       & 5,737\euro & 1,417\euro & 587\euro \\
Sum C2C Volume       & 195,070 \euro & 17,010 \euro & 16,448 \euro \\
GSCC                 & -  & -  & - \\
GIN\textbackslash GSCC    & -  & -  & - \\
GOUT\textbackslash GSCC   & -  & -  & - \\
\addlinespace
\hline\hline
Full Network & 2022 & 2023 & 2024 \\
\hline

Number of nodes      & 14,649 & 22,657 & 13,704 \\
Avg In-Degree        & 5.6 & 3.8 & 5.1 \\
Avg Out-Degree       & 5.6 & 3.8 & 5.1 \\
Avg  Degree     & 11.1 & 7.6 & 10.2 \\
Avg Volume       & 4,392 \euro & 4,500 \euro & 4,078 \euro \\
Sum Volume       & 60,193 \,k\euro & 61,668 \,k\euro & 55,745\,k\euro \\
GSCC            & 64.4\% & 35.7\% & 52.5\% \\
GIN\textbackslash GSCC     & 23.0\% & 11.8\% & 11.4\% \\
GOUT\textbackslash GSCC   & 12.3\% & 52.3\% & 35.8\% \\
\hline
\end{tabular}
}
\caption{Comparison of multilayer network metrics across 2022, 2023, 2024.}

\label{tab:multilayer_comparison}
\end{table}

The results presented in Table \ref{tab:multilayer_comparison} underscore the differentiated structural properties of the business and consumer layers, and more importantly, the critical influence of inter-layer transactions on the overall network topology and dynamics. Across the three-year observation period, the business layer exhibits relatively stable connectivity metrics. The GSCC consistently encompasses over $68\%$ of business nodes, and both the in-degree and out-degree remain balanced, suggesting a mature and well-integrated sub-network of providers. In contrast, the consumer layer shows more volatility: node counts fluctuate substantially (notably doubling in 2023), while average degrees and transaction volumes drop sharply. These observations suggest a structurally sparse and potentially fragmented layer, where intra-layer C2C exchanges are neither dense nor persistent enough to support robust component formation. Indeed, the absence of a proper GSCC and the component statistics in this layer indicate that consumers do not form large strongly connected subgraphs in isolation. However, when examining the aggregated multilayer network, a different pattern emerges. Despite the internal sparsity of the consumer layer, its role is key for the formation of the GSCC of the full network as Table \ref{tab:multilayer_comparison} shows that the GSCC of the full network outsizes by far that of the business layer in terms of number of nodes. Moreover, our multilayer representation clearly explains the decrease in the size of the GSCC observed in 2023 normalized to the number of network nodes. As this decrease is concurrent with a spike in consumer nodes and a decline in average degrees, indeed the newly added consumer nodes did not succeed in adhering to the network core. This is coherent with the observed shifts in the GOUT and GIN components. In 2023, the network exhibits a pronounced (percentage) increase in GOUT\textbackslash GSCC ($52.3\%$), coupled with a decrease in GSCC size and GIN\textbackslash GSCC. This pattern is characteristic of a structure in which many consumer nodes are reachable from the business layer but do not reciprocate transactions, forming peripheral out-components. Such asymmetry implies a breakdown in bidirectional flow, which is essential for sustaining a robust strongly connected core.

Taken together, these observations support the conclusion that inter-layer B2C and C2B transactions are not merely supplemental but are structurally integral to the connectivity and resilience of the Sarex network. In network-theoretic terms, inter-layer links act as bridges between otherwise weakly connected components, enhancing the network’s navigability, fault tolerance, and capacity for systemic coordination. The results also highlight the importance of maintaining a well-distributed pattern of inter-layer exchanges, particularly in systems where user bases fluctuate or expand rapidly. A failure to proportionally scale cross-layer connectivity—as observed in 2023—can induce fragmentation, reduce global efficiency, and weaken the system’s ability to sustain large-scale mutual reachability. 
\section{Conclusions}
\label{sec:conc}
This study has introduced a structural framework for analyzing mutual credit systems through the lens of network science, with Sardex serving as a case study. By modeling user interactions as a directed transaction network and examining its evolution over time, we provide evidence of topological patterns that are not attributable to simple topological properties alone. The observed contraction of the GSCC, the asymmetric expansion of peripheral components, and the prevalence of behavioral imitation collectively point to declining liquidity recirculation and growing structural fragmentation.
Our comparative analysis with randomized null models confirms that the Sardex network exhibits distinct macro- and meso-level properties, particularly in the imitation mechanisms, in the transactions exchanges, and in the resilience of the network thanks to the user-type heterogeneity. Moreover, the investigation of the provider role by means of the removal of their transactions reveals an underlying resilient peer-to-peer core and quite low effectiveness of provider transactions in closing loops with new users.

Several avenues for future research emerge from this study. First,  dynamic modeling of credit flows over time, potentially using agent-based or reinforcement learning models, may provide insight into how users adapt to structural incentives or constraints. Incorporating measures of economic impact or real-world business outcomes would help connect structural properties to socioeconomic performance, advancing the design of sustainable complementary currency networks. 
Finally, the results of the analyses with the condensed and multilayer graphs motivate another important research direction which is to study the role of higher-order interactions in the resilience of monetary networks—that is, interactions among groups of three or more agents that cannot be reduced to the sum of pairwise relationships. In the context of CC networks, such interactions correspond to transactions involving more than two users, where one user may compensate for the purchase of an asset by completing a cycle of transactions involving third parties.

\section{Acknowledgements}

The authors would like to thank Francesco Trudu and Paolo Piras from Sardex SpA for providing the (anonymous) data used in the analysis and Salvatore Esposito for the help in obtaining the numerical results. 

\bibliographystyle{IEEEtran} 
\bibliography{biblio}

\appendices
\section{Numerical Data} 
\subsection{Incoming and outgoing users} \label{app1}

Nodes that left the market in 2022 participated in both buying and selling, handling about 10\% of outgoing transactions but less than 5\% of incoming transactions and less than 5\% of total annual trading volume  (Table~\ref{tab:nodeExit}). Differently, the majority of nodes that exited the market in 2023 were primarily sellers with no purchasing activity, but still contributing to less than 5\% of transactions and under 6\% of the total volume (Table~\ref{tab:nodeExit}). Instead, users who remained active from 2022 to 2024 handled over $80\%$ of transactions and traded volume (Table~\ref{tab:nodeActiv}). 


\begin{table}[!th]
    \centering
    \resizebox{0.48\textwidth}{!}{
    \begin{tabular}{rrrrrr}
    \hline
$2022$    & $\sum_{i} \lambda_i$&$\sum_{i \in \mathcal{N}} \lambda_i \theta_i^{\text{out}}$&$\sum_{i \in \mathcal{N}} \lambda_i\theta_i^{\text{in}}$&$\sum_{i \in \mathcal{N}} \lambda_iv_i^{\text{out}}$&$\sum_{i \in \mathcal{N}} \lambda_iv_i^{\text{in}}$\\\hline
$\theta_i^{\text{in}} > 0 \; \wedge \; 
 \theta_i^{\text{out}} > 0 $ &    2,010&39,751&15,553&1,725\,k\euro&2,134\,k\euro\\
$\theta_i^{\text{in}} > 0 \; \wedge \; 
 \theta_i^{\text{out}} = 0 $&1,034&-&3,839& - &784\,k\euro\\
$\theta_i^{\text{in}} = 0 \; \wedge \; 
 \theta_i^{\text{out}} > 0 $&1,834&5,276& - &275\,k\euro& - \\\hline
\multirow{2}{*}{total}
 &4,878&45,007&19,392&2,000\,k\euro&2,917\,k\euro\\
&($33.3\%$)&($11.1\%$)&($4.8\%$)&($3.3\%$)&($4.8\%$)\\ \hline\hline
$2023$    & $\sum_{i} \lambda_i$&$\sum_{i \in \mathcal{N}} \lambda_i \theta_i^{\text{out}}$&$\sum_{i \in \mathcal{N}} \lambda_i \theta_i^{\text{in}}$&$\sum_{i \in \mathcal{N}} \lambda_i v_i^{\text{out}}$&$\sum_{i \in \mathcal{N}} \lambda_i v_i^{\text{in}}$\\\hline
$\theta_i^{\text{in}} > 0 \; \wedge \;  \theta_i^{\text{out}} > 0 $ &1,204&11,504&13,423&2,073\,k\euro&2,594\,k\euro\\
$\theta_i^{\text{in}} > 0 \; \wedge \; \theta_i^{\text{out}} = 0 $&10,946& - &13,414& - &1,013\,k\euro\\
$\theta_i^{\text{in}} = 0 \; \wedge \;  \theta_i^{\text{out}} > 0 $&1,271&3,170& -& 303\,k\euro& - \\\hline
\multirow{2}{*}{total}  &13,421&14,724&26,837&2,376\,k\euro&3,607\,k\euro\\
&($59.2\%$)&($4.2\%$)&($7.6\%$)&($3.9\%$)&($5.9\%$)\\\hline
    \end{tabular}}
    \caption{Transactions and volumes of the nodes inactive or exiting the circuit during $2022$ and $2023$; the logic variable $\lambda_i \in \{0,1\}$ is given by the condition in the first column.}
    \label{tab:nodeExit}
\end{table}

\begin{table}[!th]
    \centering
    \resizebox{0.48\textwidth}{!}{
    \begin{tabular}{rrrrrr}
\hline 2022    & $\sum_{i} \lambda_i$&$\sum_{i \in \mathcal{N}} \lambda_i \theta_i^{\text{out}}$&$\sum_{i \in \mathcal{N}} \lambda_i \theta_i^{\text{in}}$&$\sum_{i \in \mathcal{N}} \lambda_i v_i^{\text{out}}$&$\sum_{i \in \mathcal{N}} \lambda_i v_i^{\text{in}}$\\\hline
$\theta_i^{\text{in}} > 0 \; \wedge \; 
 \theta_i^{\text{out}} > 0 $ &6,097&313,564&348,380&53,451\,k\euro&51,392\,k\euro\\
$\theta_i^{\text{in}} > 0 \; \wedge \; \theta_i^{\text{out}} = 0 $&326& - &2,297& - &480\,k\euro\\
$\theta_i^{\text{in}} = 0 \; \wedge \;  \theta_i^{\text{out}} > 0 $&958&7,430& - &472\,k\euro& - \\
\hline
\multirow{2}{*}{total}  &7,381&320,994&350,677&53,923\,k\euro&51,873\,k\euro\\
&(50.4\%)&(79.4\%)&(86.8\%)&(89.6\%)&(86.2\%)\\\hline \hline
2023    & $\sum_{i} \lambda_i$&$\sum_{i \in \mathcal{N}} \lambda_i \theta_i^{\text{out}}$&$\sum_{i \in \mathcal{N}} \lambda_i \theta_i^{\text{in}}$&$\sum_{i \in \mathcal{N}} \lambda_i v_i^{\text{out}}$&$\sum_{i \in \mathcal{N}} \lambda_i v_i^{\text{in}}$\\\hline
$\theta_i^{\text{in}} > 0 \; \wedge \; 
 \theta_i^{\text{out}} > 0 $ &6,010&313,962&305,945&54,190\,k\euro&53,439\,k\euro\\
$\theta_i^{\text{in}} > 0 \; \wedge \; \theta_i^{\text{out}} = 0 $&453& - &3,456& - &876\,k\euro\\
$\theta_i^{\text{in}} = 0 \; \wedge \;  \theta_i^{\text{out}} > 0 $&918&11,556& - &233\,k\euro& - \\
\hline
\multirow{2}{*}{total}  &7,381&325,518&309,401&54,424\,k\euro&54,315\,k\euro\\
&(32.6\%)&(91.9\%)&(87.3\%)&(88.2\%)&(88.1\%)\\\hline \hline
2024    & $\sum_{i} \lambda_i$&$\sum_{i \in \mathcal{N}} \lambda_i \theta_i^{\text{out}}$&$\sum_{i \in \mathcal{N}} \lambda_i \theta_i^{\text{in}}$&$\sum_{i \in \mathcal{N}} \lambda_i v_i^{\text{out}}$&$\sum_{i \in \mathcal{N}} \lambda_i v_i^{\text{in}}$\\\hline
$\theta_i^{\text{in}} > 0 \; \wedge \; 
 \theta_i^{\text{out}} > 0 $ &5,346&279,546&263,482&46,339\,k\euro&45,572\,k\euro\\
$\theta_i^{\text{in}} > 0 \; \wedge \; \theta_i^{\text{out}} = 0 $&1,068& - &6,889& - &1,212\,k\euro\\
$\theta_i^{\text{in}} = 0 \; \wedge \;  \theta_i^{\text{out}} > 0 $&967&8,154& - &256\,k\euro& - \\
\hline
\multirow{2}{*}{total}  &7,381&287,700&270,371&46,595\,k\euro&46,784\,k\euro\\
&(53.9\%)&(88.5\%)&(83.2\%)&(83.6\%)&(83.9\%)\\\hline 
    \end{tabular}}
    \caption{Transactions and volumes per year of the nodes which remain active for all the three years of analysis (2022, 2023, 2024); the logic variable $\lambda_i \in \{0,1\}$ is given by the condition in the first column.}
    \label{tab:nodeActiv}
\end{table}


The total number of Sardex transactions is $403,995$ in 2022 for a volume of $60,193$\,k\euro, $354,317$ in 2023 for a volume of $61,668$\,k\euro, and $325,078$ in 2024 for a volume of $55,745$\,k\euro. In Table \ref{tab:i_numeri}, these quantities are observed by grouping according to the degrees of in and out of the various users.

\begin{table}[!th]
    \centering
    \resizebox{0.48\textwidth}{!}{
    \begin{tabular}{rrrrr}
    \hline
    &  $\sum_{i} \lambda_i$& $\sum_{i,j\in \mathcal{N}} \lambda_i e_{ij}$ & $\sum_{i,j\in \mathcal{N}} \lambda_i w_{ij}$ & $\sum_{i,j\in \mathcal{N}} \lambda_i \delta_{ij}$ \\\hline
\textbf{2022} &\textbf{14,649} & \textbf{403,995} &\textbf{60,193\,k\euro }& \textbf{81,328} \\\hline
$\theta_i^{\text{in}} \geq 1 \; \wedge \; 
 \theta_i^{\text{out}} = 0 $&11.5\%& 1.9\% & 2.8\% &4.3\% \\
$\theta_i^{\text{in}} = 0 \; \wedge \; 
 \theta_i^{\text{out}} \geq 1 $  &23.1\% & 3.9\% & 1.5\% &5.4\%\\
$\theta_i^{\text{in}} \geq 1 \; \wedge \; 
 \theta_i^{\text{out}} \geq 1 $ &65.4\% & 94.2\% & 95.7\% &90.3\% \\\hline
$\theta_i^{\text{in}} = 0 \; \wedge \; 
 \theta_i^{\text{out}} = 1 $&12.3\%& 0.5\% & 0.6\% &2.2\% \\
 $\theta_i^{\text{in}} > 0 \; \wedge \; 
 \theta_i^{\text{out}} = 1 $ &9.7\% & $< 0.1$\% & 0.2\% & $< 0.1$\%\\\hline
$\theta_i^{\text{in}} = 1 \; \wedge \; 
 \theta_i^{\text{out}} = 0 $ &5.9\% & 0.2\% & 0.7\% & 1.1\% \\
$\theta_i^{\text{in}} = 1 \; \wedge \; 
 \theta_i^{\text{out}} > 0 $&10.1\% & $< 0.1$\% & $< 0.1$\% &$< 0.1$\% \\\hline
$\theta_i^{\text{in}} > 1 \; \wedge \; 
 \theta_i^{\text{out}} > 1$ &48.1\% & 87.9\% & 87.4\% & 79.4\% \\\hline\hline
\textbf{2023}&\textbf{22,657} & \textbf{354,317} &\textbf{61,668\,k\euro}&\textbf{86,443} \\\hline
$\theta_i^{\text{in}} \geq 1 \; \wedge \; 
 \theta_i^{\text{out}} = 0 $&51.9\% & 5.0\% & 3.6\% &1.6\% \\
$\theta_i^{\text{in}} = 0 \; \wedge \; 
 \theta_i^{\text{out}} \geq 1 $  &11.9\% & 4.6\% & 1.2\% &4.1\% \\
$\theta_i^{\text{in}} \geq 1 \; \wedge \; 
 \theta_i^{\text{out}} \geq 1 $ &36.3\% & 90.5\% & 95.2\%& 80.0\% \\\hline
 $\theta_i^{\text{in}} = 0 \; \wedge \; 
 \theta_i^{\text{out}} = 1 $ &6.0\% & 0.4\% & 0.5\% &1.6\% \\
 $\theta_i^{\text{in}} > 0 \; \wedge \; 
 \theta_i^{\text{out}} = 1 $ &6.2\% & $< 0.1$\% & $< 0.1$\% &$< 0.1$\% \\\hline
$\theta_i^{\text{in}} = 1 \; \wedge \; 
 \theta_i^{\text{out}} = 0 $ &48.0\% & 3.1\% & 1.1\% &12.6\% \\
$\theta_i^{\text{in}} = 1 \; \wedge \; 
 \theta_i^{\text{out}} > 0 $&6.1\% & $< 0.1$\% & $< 0.1$\% &$< 0.1$\% \\\hline
$\theta_i^{\text{in}} > 1 \; \wedge \; 
 \theta_i^{\text{out}} > 1 $&25.8\% & 82.5\% & 88.9\%&70.0\% \\\hline\hline
\textbf{2024}&\textbf{13,704} & \textbf{325,078} & \textbf{55,745\,k\euro} &\textbf{69,986} \\\hline
$\theta_i^{\text{in}} \geq 1 \; \wedge \; 
 \theta_i^{\text{out}} = 0 $&35.1\% & 3.9\% & 4.2\% &10.2\% \\
$\theta_i^{\text{in}} = 0 \; \wedge \; 
 \theta_i^{\text{out}} \geq 1 $ &11.5\% & 3.4\% & 1.4\% &3.3\% \\
$\theta_i^{\text{in}} \geq 1 \; \wedge \; 
 \theta_i^{\text{out}} \geq 1 $&53.4\% & 92.6\%& 94.5\% &86.5\% \\ \hline
$\theta_i^{\text{in}} = 0 \; \wedge \; 
 \theta_i^{\text{out}} = 1 $ &3.9\% & 0.2\% & 0.4\% &0.8\% \\
$\theta_i^{\text{in}} > 0 \; \wedge \; 
 \theta_i^{\text{out}} = 1 $&7.9\% & $< 0.1$\% & $< 0.1$\% &$< 0.1$\% \\ \hline
$\theta_i^{\text{in}} = 1 \; \wedge \; 
 \theta_i^{\text{out}} = 0 $ &26.0\% & 1.1\% & 1.1\% &5.1\% \\
$\theta_i^{\text{in}} = 1 \; \wedge \; 
 \theta_i^{\text{out}} > 0 $&7.7\% & $< 0.1$\% & $< 0.1$\% &$< 0.1$\% \\ \hline
$\theta_i^{\text{in}} > 1 \; \wedge \; 
 \theta_i^{\text{out}} > 1 $&40.2\% & 86.1\% & 86.6\% &77.0\% \\\hline

    \end{tabular}}
    \caption{The Sardex circuit data for different ranges of transactions (2022, 2023, 2024); the logic variable $\lambda_i \in \{0,1\}$ is given by the condition in the first column.}
    \label{tab:i_numeri}
\end{table}

To assess whether the transactional behavior of newly joined users significantly deviates from that of the more established user base, we analyze average and standard deviation of ingoing and outgoing transaction volumes for new users introduced to the Sardex network in 2023 against those of the overall user population during the same year, in Table \ref{tab:volume_comparison}. On average, new users exhibit slightly lower transaction volumes than the general user population: the mean ingoing volume for new users is approximately 2,618\euro, compared to 2,778\euro~for all users; similarly, the mean outgoing volume is 2,648\euro\, for new users versus 2,762\euro\, for all users. More notably, the standard deviations for both ingoing and outgoing volumes are substantially lower for new users (13,124\euro\, and 14,121\euro, respectively) than for the overall user base (18,651\euro\, and 21,415\euro).  Overall, these results suggest that while new users integrate into the network with meaningful transaction activity, they tend to do so with lower variance and slightly lower volumes. This homogeneity may be indicative of standardized entry-level participation of new users. 

\begin{table}[ht]
\centering
\resizebox{\columnwidth}{!}{%
\begin{tabular}{lcccc}
\hline
User Group & Mean Ingoing & Std Ingoing & Mean Outgoing & Std Outgoing \\
\hline
New Users (2023) & 2,618 & 13,124 & 2,648 & 14,121 \\
All Users (2023) & 2,778 & 18,651 & 2,762 & 21,415\\
\hline
\end{tabular}%
}
\caption{Comparison of transaction volumes between new users in 2023 and all the remaining users in 2023.}
\label{tab:volume_comparison}
\end{table}

\subsection{Transactions for different type users} \label{app2}

In Tables \ref{tab:tabtypeB}--\ref{tab:tabBilanci}, the transactions, their volumes, and some quantities are analyzed in relation to the type of user. In these tables B stands for Business, C for Consumer, E for Employee, and P for provider (Gestore).

\begin{table}[!th]
    \centering
    \resizebox{0.45\textwidth}{!}{
    \begin{tabular}{rrrp{0.3cm}rrr}
    \hline
     Type&Years& \multicolumn{3}{c}{$N^{\text{out}} \rightarrow N^{\text{in}}$} & \multicolumn{1}{c}{$\sum_{i,j\in \mathcal{N}} e_{ij}$} &\multicolumn{1}{c}{$\sum_{i,j\in \mathcal{N}} w_{ij}$} \\ \hline 
\multirow{3}{*}{$B \rightarrow B $}     &2022&4,178&$\to$&	4,353&	164,478&	42,544\,k\euro\\
&2023&	3,901	 &$\to$ &	4,081	&	150,438	&	43,935\,k\euro	\\
&2024&3,467&$\to$&3,633&	133,518&	39,831\,k\euro\\\hline
\multirow{3}{*}{$ B \rightarrow C$}	&2022&475&$\to$ &	3,763&	15,363&	254\,k\euro\\
&2023&	583	 &$\to$ &	2,585	&	10,014	&	280\,k\euro	\\
&2024&520&$\to$ &2,281&	9,633&	251\,k\euro\\\hline
\multirow{3}{*}{$ B \rightarrow E$}	&2022&2,066&$\to$ &	2,396&	98,318&	5,950\,k\euro\\
&2023&	2,018	 &$\to$ &	2,436	&	99,607	&	6,403\,k\euro	\\
&2024&1,822&$\to$ &	2,609&	99,886&	6,262\,k\euro \\ \hline
\multirow{3}{*}{$ B \rightarrow P$}	&2022&380&$\to$ &	2&	953&2,554\,k\euro\\
&2023&	394	 &$\to$ &	4	&	990	&	1,674\,k\euro\\
&2024&354&$\to$ &	2&	933&	871\,k\euro \\ \hline
 \end{tabular}
 }
\caption{Sardex transaction data for B users (2022, 2023, 2024).}
    \label{tab:tabtypeB}
 \end{table}

\begin{table}[!th]
    \centering
    \resizebox{0.45\textwidth}{!}{
    \begin{tabular}{rrrp{0.3cm}rrr}\hline
     Type&Years& \multicolumn{3}{c}{$N^{\text{out}} \rightarrow N^{\text{in}}$} & \multicolumn{1}{c}{$\sum_{i,j\in \mathcal{N}} e_{ij}$} &\multicolumn{1}{c}{$\sum_{i,j\in \mathcal{N}} w_{ij}$} \\ \hline 
\multirow{3}{*}{$ C \rightarrow B$} &2022&6,392&$\to$ &	375&	106,596	&352\,k\euro\\
&2023&	4,000	 &$\to$ &	330	&	63,125	&	323\,k\euro	\\
&2024&2,953 &$\to$ &	249&	59,442&	175\,k\euro \\ \hline
\multirow{3}{*}{$ C \rightarrow C$}	&2022&0 &$\to$ &0&	0&0\euro\\
&2023&	0	 &$\to$ &	0	&	0	&	0\euro	\\
&2024&0&$\to$ &0&0&0\euro \\ \hline
\multirow{3}{*}{$ C \rightarrow E$}	&2022&1 &$\to$ &	1&	3&	800\euro\\
&2023&	1	 &$\to$ &	1	&	2	&	350\euro	\\
&2024&0&$\to$ &0&0&0\euro \\ \hline
\multirow{3}{*}{$ C \rightarrow P$}	&2022&2&$\to$ &	1&	2&	1\,k\euro\\
&2023&	455	 &$\to$ &	1	&	456	&	14\,k\euro	\\
&2024&83&$\to$ &	1&	87&	2\,k\euro \\ \hline
 \end{tabular}
 }
\caption{Sardex transaction data for C users  (2022, 2023, 2024).}
    \label{tab:tabtypeC}
 \end{table}

\begin{table}[!th]
    \centering
    \resizebox{0.45\textwidth}{!}{
    \begin{tabular}{rrrp{0.3cm}rrr}
    \hline
     Type&Years& \multicolumn{3}{c}{$N^{\text{out}} \rightarrow N^{\text{in}}$} &\multicolumn{1}{c}{$\sum_{i,j\in \mathcal{N}} e_{ij}$} &\multicolumn{1}{c}{$\sum_{i,j\in \mathcal{N}} w_{ij}$} \\ \hline 
\multirow{3}{*}{$ E \rightarrow B$}	&2022&2,132&$\to$ &	845&	14,362&	6,333\,k\euro\\
&2023&	2,431	 &$\to$ &	947	&	15,314	&	6,698\,k\euro	\\
&2024&2,167&$\to$ &	849&	14,650&	6,183\,k\euro \\ \hline
\multirow{3}{*}{$ E \rightarrow C$}	&2022&1	 &$\to$ &	1	&	1	&101\euro	\\
&2023&	0	 &$\to$ &	0	&	0	&	0\euro	\\
&2024&0&$\to$ &0&0&0\euro \\ \hline
\multirow{3}{*}{$ E \rightarrow E$}	&2022&29&$\to$ &29&	34&194\,k\euro\\
&2023&	10	 &$\to$ &	11	&	13	&	17\,k\euro	\\
&2024&24&$\to$ &	28&	34&	16\,k\euro \\ \hline
\multirow{3}{*}{$ E \rightarrow P$}	&2022&73&$\to$ &	2&	236&	120\,k\euro\\
&2023&	67	 &$\to$ &	2	&	237	&	100\,k\euro	\\
&2024&65&$\to$ &	2&	284&	136\,k\euro \\ \hline
 \end{tabular}
 }
\caption{Sardex transaction data for E users.}
    \label{tab:tabtypeE}
 \end{table}

\begin{table}[!th]
    \centering
    \resizebox{0.45\textwidth}{!}{
    \begin{tabular}{rrrp{0.3cm}rrr}
    \hline
     Type&Years& \multicolumn{3}{c}{$N^{\text{out}} \rightarrow N^{\text{in}}$} & \multicolumn{1}{c}{$\sum_{i,j\in \mathcal{N}} e_{ij}$} &\multicolumn{1}{c}{$\sum_{i,j\in \mathcal{N}} w_{ij}$}\\ \hline 
\multirow{3}{*}{$ P\rightarrow B $}	&2022&2&$\to$ &	2,498&	3,623&	1,851\,k\euro\\
&2023&	2	 &$\to$ &	2,907	&	4,176	&	2,130\,k\euro	\\
&2024&2&$\to$ &	2,543&	3,816&	1,964\,k\euro\\ \hline
\multirow{3}{*}{$ P \rightarrow C$}	&2022&1&$\to$ &	1&	1&	500\euro\\
&2023&	1	 &$\to$ &	9,922	&	9,922	&	79\,k\euro	\\
&2024&1	 &$\to$ & 2,772&	2,772&	39\,k\euro \\ \hline
\multirow{3}{*}{$ P \rightarrow E$}	&2022&2&$\to$ &	18&	23&	22\,k\euro\\
&2023&	2	 &$\to$ &	16	&	23	&	15\,k\euro	\\
&2024&2 &$\to$ &	18&	23&	15\,k\euro \\ \hline
\multirow{3}{*}{$ P \rightarrow G$}	&2022&2&$\to$ &	2&	2&	16\,k\euro\\
&2023&	0	 &$\to$ &	0	&	0	&	0\euro	\\
&2024& 0&$\to$ &0&0&0\euro \\ \hline 
 \end{tabular}
 }
\caption{Sardex transaction data for P users  (2022, 2023, 2024).}
    \label{tab:tabtypeG}
 \end{table}

\begin{table}[!th]
\centering
\resizebox{0.45\textwidth}{!}{
\begin{tabular}{rrrrrrr}\\
\hline
2022& [0, 1\euro] & (1\euro, 10\euro]& (10\euro, 100\euro]& (100\euro, 1\,k\euro]& (1\,k\euro, 10\,k\euro]& > 10\,k\euro\\ \hline
 B &<1\%&14.5\%&59.8\%&22.4\%&2.8\%&<1\%\\
 C &65.0\%&33.1\%&1.4\%&<1\%&<1\%&<1\%\\
 E &<1\%&<1\%&23.0\%&68.4\%&6.6\%&<1\%\\
 P &2.6\%&2.8\%&20.4\%&65.1\%&8.8\%&<1\%\\ \hline \hline
2023& [0, 1\euro] & (1\euro, 10\euro]& (10\euro, 100\euro]& (100\euro, 1\,k\euro]& (1\,k\euro, 10\,k\euro]& > 10\,k\euro\\ \hline
 B &<1\%&13.4\%&59.4\%&24.7\%&3.1\%&<1\%\\
 C &54.4\%&42.9\%&1.6\%&1.1\%&<1\%&<1\%\\
 E &1.6\%&2.0\%&21.1\%&68.4\%&6.8\%&<1\%\\
 P &13.7\%&42.6\%&20.5\%&20.4\%&2.7\%&<1\%\\ \hline \hline
2024& [0, 1\euro] & (1\euro, 10\euro]& (10\euro, 100\euro]& (100\euro, 1\,k\euro]& (1\,k\euro, 10\,k\euro]& > 10\,k\euro\\ \hline
 B &<1\%&14.8\%&59.0\%&22.7\%&3.0\%&<1\%\\
 C &55.6\%&42.8\%&1.3\%&<1\%&<1\%&<1\%\\
 E &1.4\%&2.0\%&24.6\%&66.1\%&5.7\%&<1\%\\
 P &8.4\%&22.3\%&26.8\%&37.0\%&5.3\%&<1\%\\ \hline 
\end{tabular}
 }
\caption{Distribution of the number of transactions (with respect to the total number of transactions) for different ranges of monetary values and user type (2022, 2023, 2024).}
    \label{tab:tabVolumi}
\end{table}

\begin{table}[!th]
\centering
\begin{tabular}{rrrrr} 
\hline
Year&Type& $\pm 5$\euro& $\pm 50$\euro& $> 0$\\ \hline
\multirow{4}{*}{2022} &B&1.8\%&4.8\%&54.6\%\\
& C&31.8\%&74.4\%&38.2\%\\
& E&8.7\%&28.2\%&43.3\%\\
& P&<1\%&	<1\%&14.3\% \\ \hline \hline
Year&Type& $\pm 5$\euro& $\pm 50$\euro& $> 0$\\ \hline
\multirow{4}{*}{2023}& B&1.7\%&4.4\%&58.3\%\\
& C&22.4\%&76.8\%&77.0\%\\
& E&6.7\%&22.6\%&	39.5\%\\
& P&<1\%&	<1\%	&21.4\%\\\hline \hline
Year&Type& $\pm 5$\euro& $\pm 50$\euro& $> 0$\\ \hline
\multirow{4}{*}{2024}& B&1.9\%&4.7\%&57.54\%\\
& C&42.3\%&69.8\%&66.8\%\\
& E&7.4\%\%&24.2\%&54.7\%\\
& P&14.3\%&	14.3\%&<1 \%\\ \hline 
\end{tabular}
\caption{Percentage of users distinguished by type with balance included in the intervals $\pm 5$\euro and $\pm 50$\euro and with positive balance (2022, 2023, 2024).}
    \label{tab:tabBilanci}
\end{table}


\nocite{*}
\subsection{Geolocal clustering}

\begin{figure}
\centering
\includegraphics[width=0.15\textwidth]{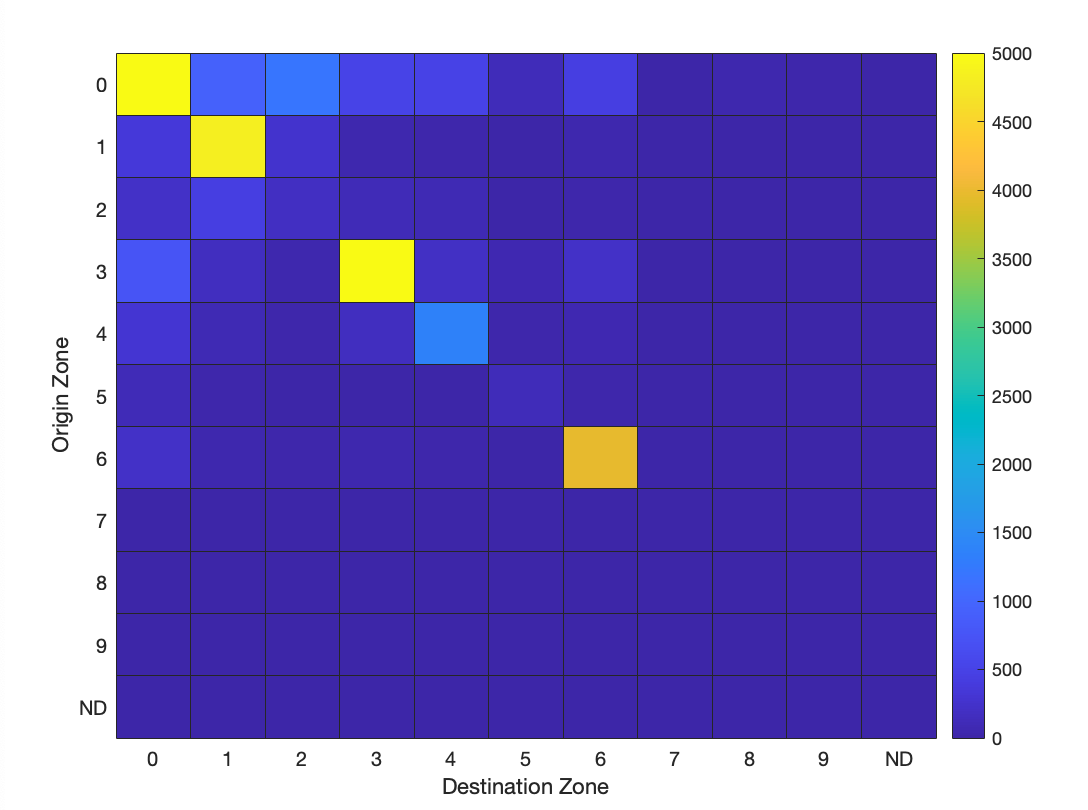}
\includegraphics[trim = {0mm 0mm 0mm 12mm},clip,width=0.157\textwidth]{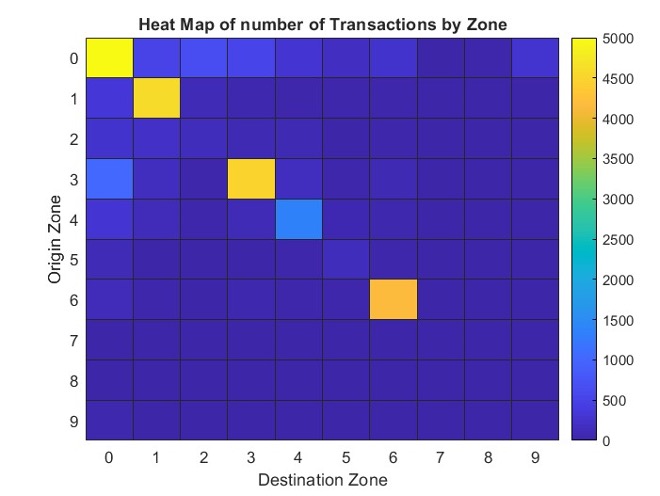}
\includegraphics[width=0.14\textwidth]{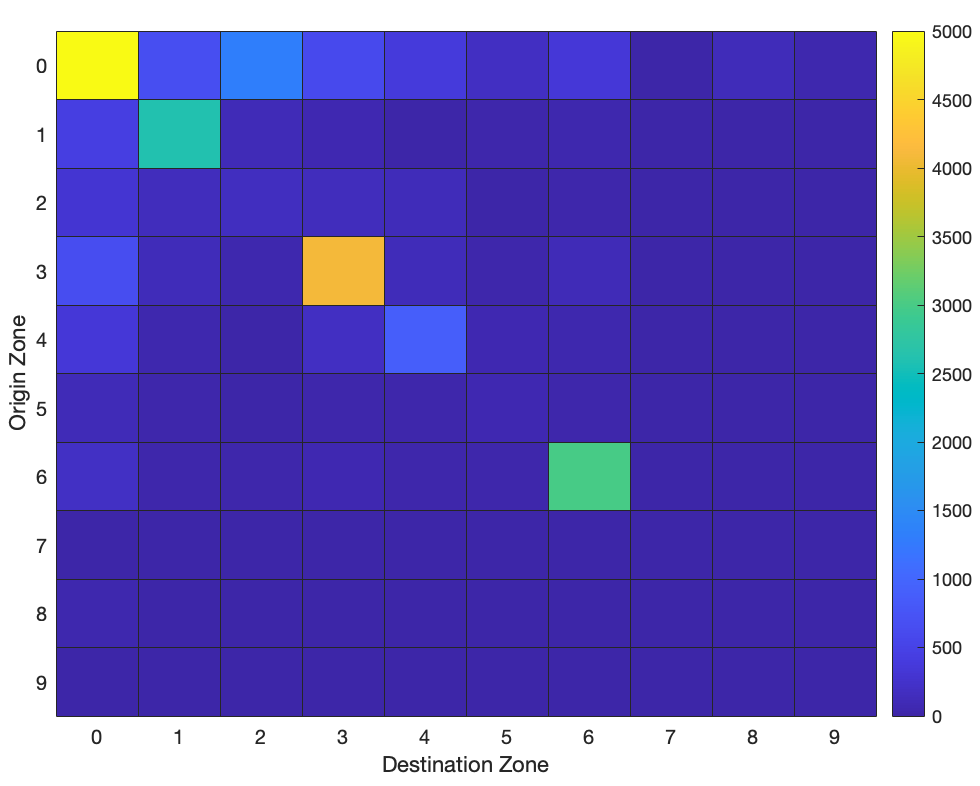}
\caption{Heat maps of the mean number of Sardex credit transactions among Italian zones for years 2022 (left), 2023 (center), and 2024 (right), respectively.} 
\label{fig:heatmaps_number_zone}
\end{figure}

\begin{figure}[!h]
\centering
\includegraphics[width=0.15\textwidth]{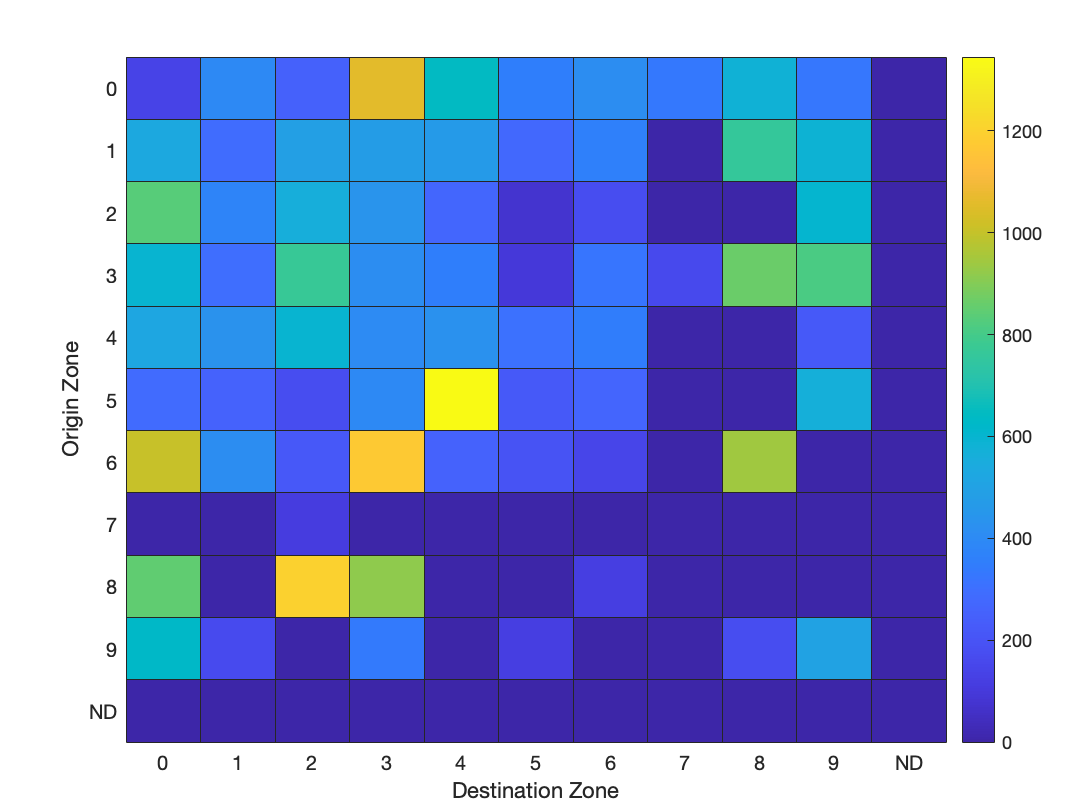}
\includegraphics[trim = {0mm 0mm 0mm 12mm},clip,width=0.157\textwidth]{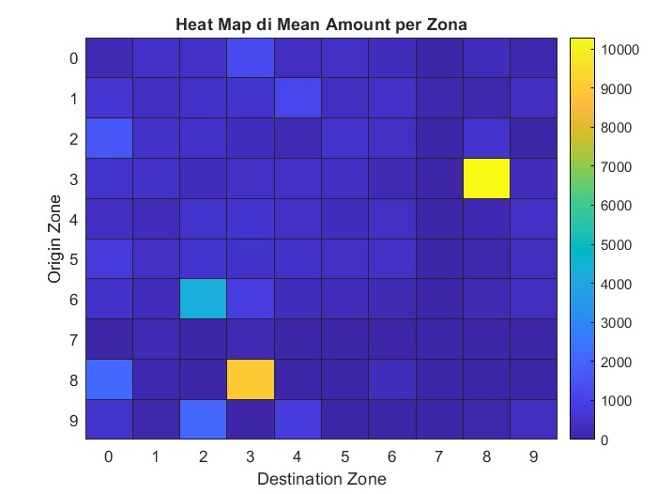}
\includegraphics[width=0.14\textwidth]{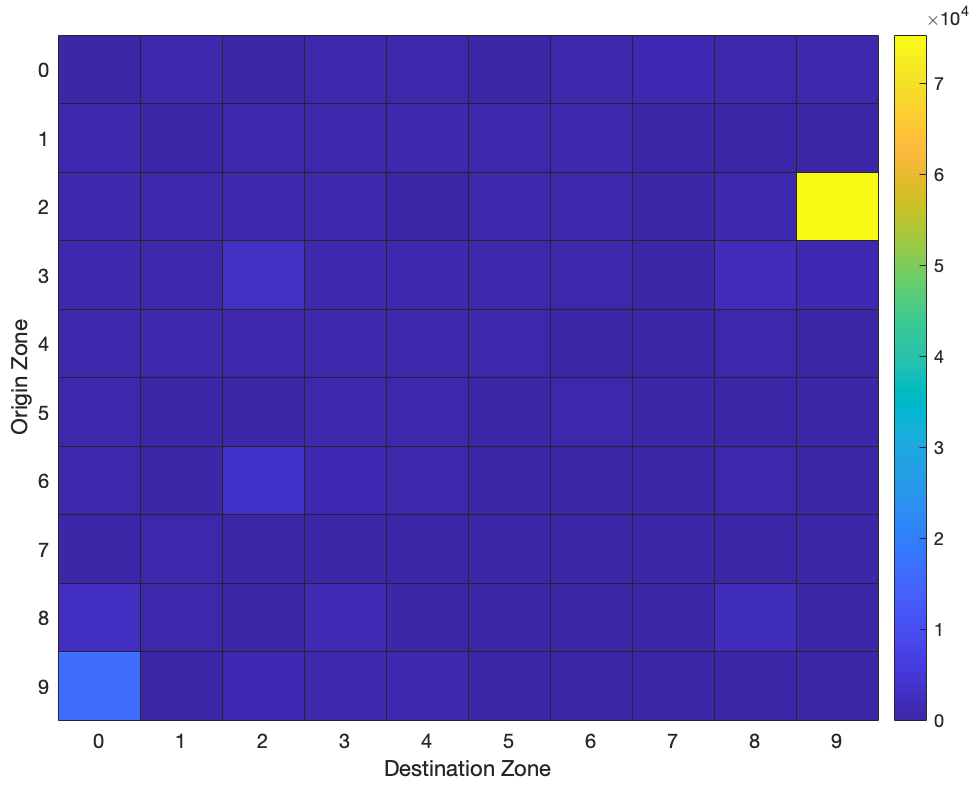}
\caption{Heat maps of the mean volume of Sardex credit exchanged among Italian zones for years 2022 (left), 2023 (center), and 2024 (right), respectively.} 
\label{fig:heatmaps_amount_zone}
\end{figure}

To explore how the geolocalization of users influences the transactions' schemes of such network, for all the years under consideration, we aggregated the users in 10 zones, according to the first number of the Italian postal code (from 0 to 9). This type of clustering highlights the zones that exchange the highest number of transactions, see Figure~\ref{fig:heatmaps_number_zone}, and the highest amount of transactions, see Figure~\ref{fig:heatmaps_amount_zone}. 
In particular, when considering the number of transactions exchanged among zones, across all the years taken in account, the intra-zone transactions are more frequent than the inter-zone ones, with the only exception of zone 0 that contains Sardinia, the core of Sardex network, that receives a substantial number of transaction (about 1500) from zone 3 in North-East of Italy, and it is the origin of quite exchanges with Northern Italy, with marginal changes from year 2022 to 2023 and a slight decrease in intra-zone transactions in favor of an increase of exchanges from Sardinia in 2024. These considerations do not hold true when considering the volume of Sardex credit exchanged rather than the number of transactions: indeed, in this case, we see that the e-money is exchanged mostly between zone 3 and zone 8 (slightly more from zone 3 to zone 8 than vice versa) and from zone 6 to zone 2.

\subsection{Business sector clustering}

\begin{figure}[!h]
\centering
\includegraphics[trim = {4mm 1mm 1mm 1mm}, clip, width=0.15\textwidth]{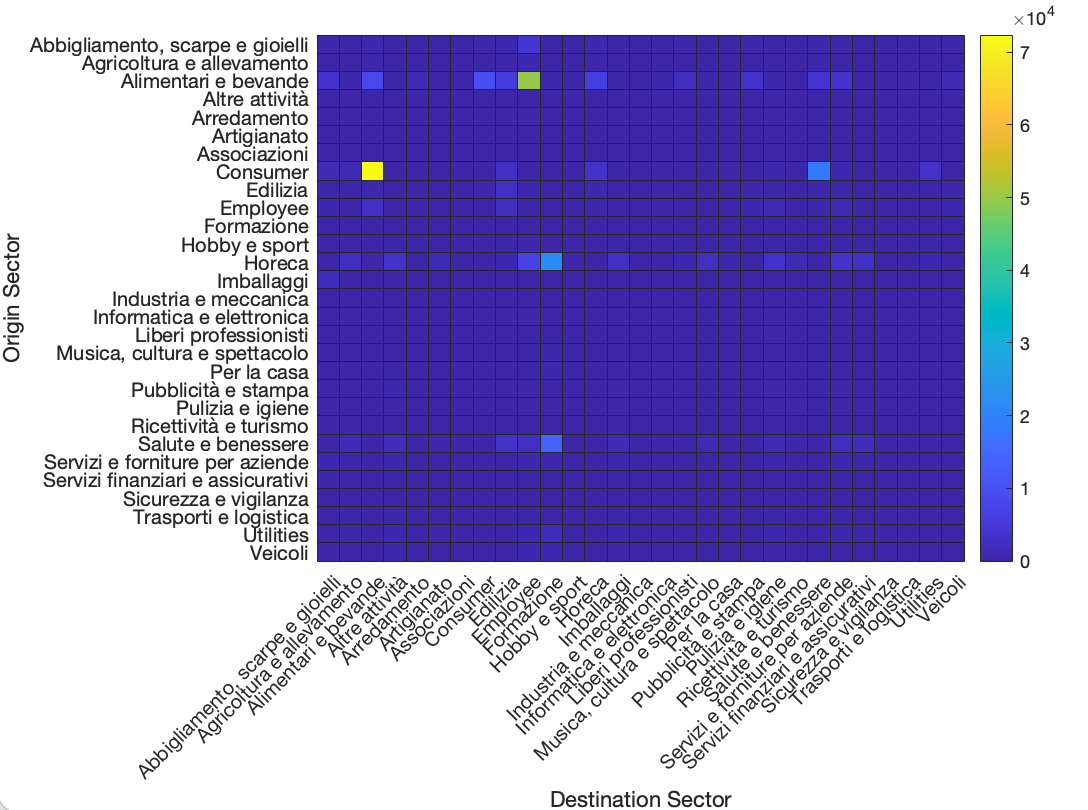}
\includegraphics[trim = {1mm 1mm 1mm 1mm}, clip, width=0.15\textwidth]{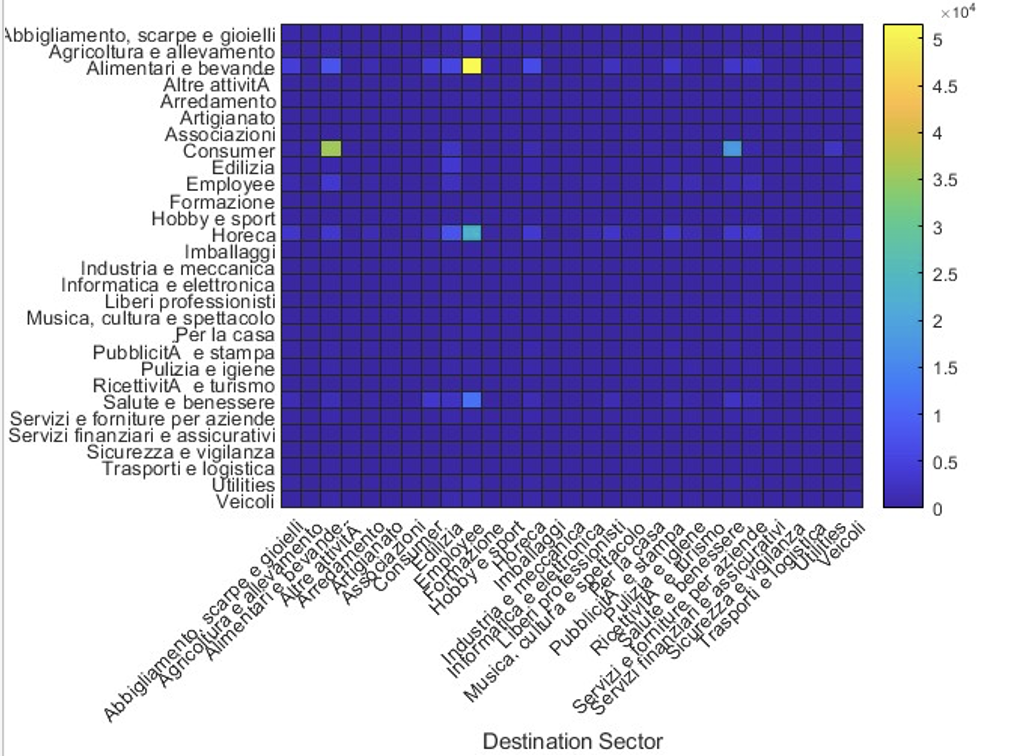}
\includegraphics[trim = {5mm 1mm 1mm 1mm}, clip, width=0.15\textwidth]{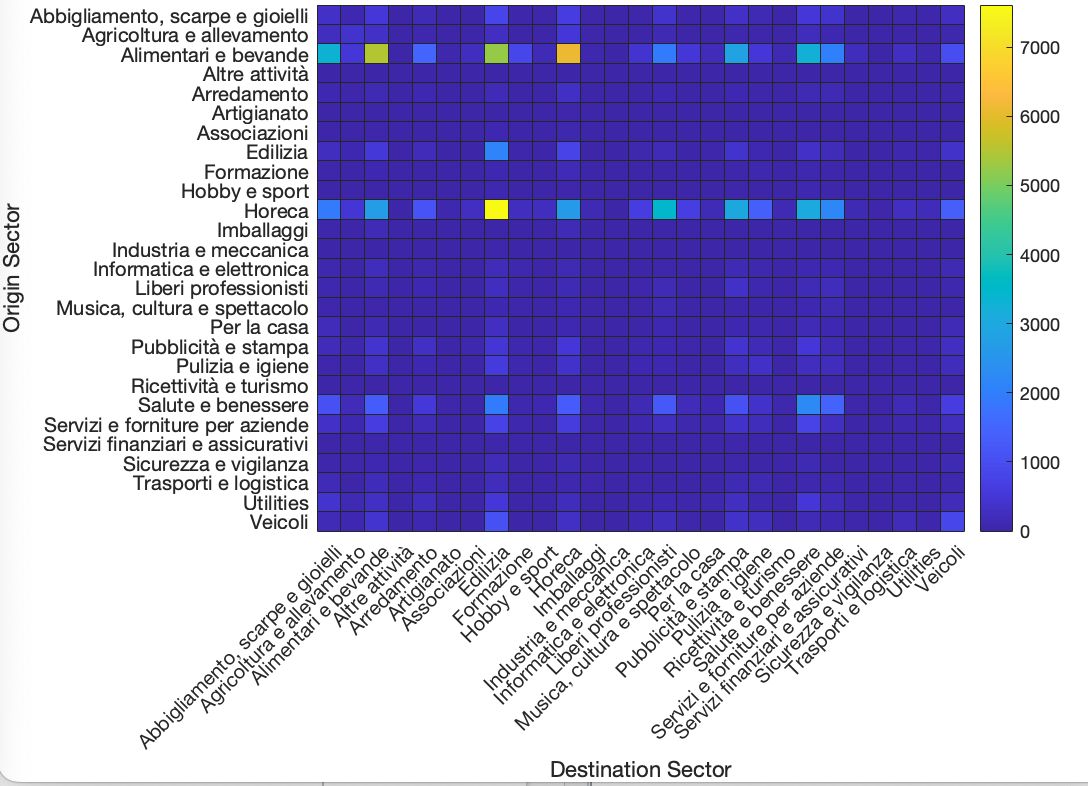}
\caption{Heat maps of the mean number of Sardex credit transactions among the Sardex business sector for years 2022 (left), 2023 (center), and 2024 (right), respectively. Each heat map visualizes the number of exchanges between different origin and destination sectors, with more frequent transactions represented by warmer colors.} 
\label{fig:heatmaps_number_sector}
\end{figure}

\begin{figure}[!h]
\centering
\includegraphics[trim = {4mm 0mm 1mm 1mm}, clip, width=0.15\textwidth]{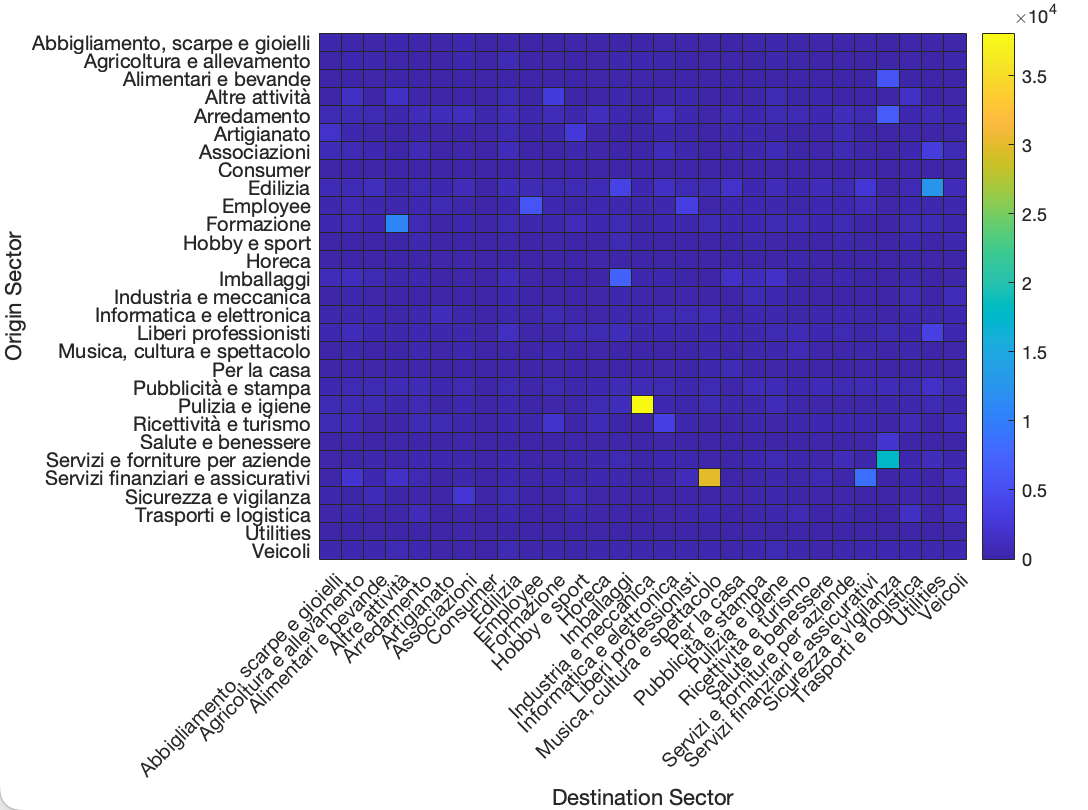}
\includegraphics[trim = {5.5mm 0mm 1mm 1mm},clip, width=0.15\textwidth]{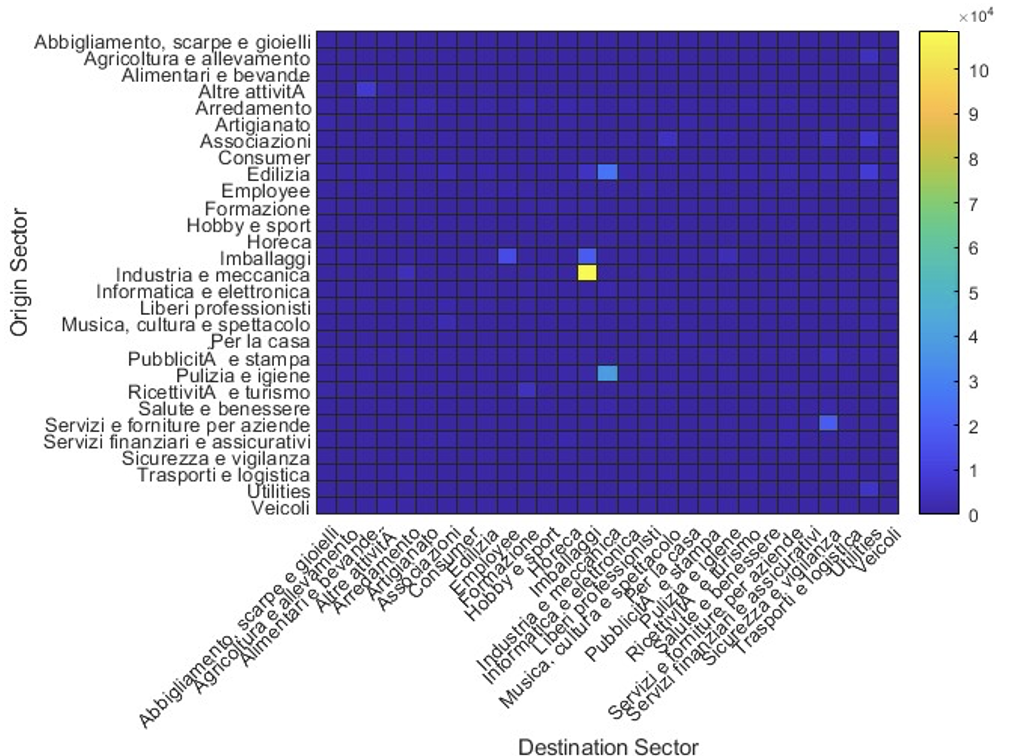}
\includegraphics[trim = {5mm 0mm 1mm 1mm},clip, width=0.15\textwidth]{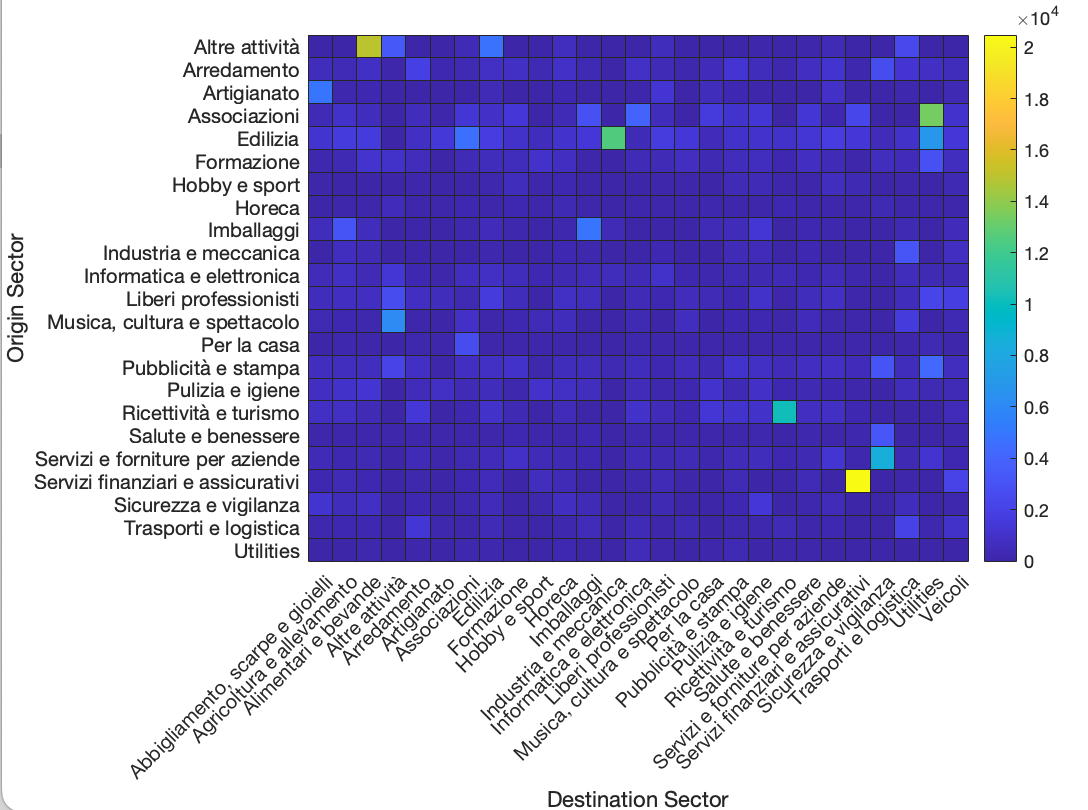}
\caption{Heat maps of the mean volume of Sardex credit transactions among the Sardex business sectors for years 2022 (left), 2023 (center), and  2024 (right), respectively. Each heat map visualizes the intensity of exchanges between different origin and destination sectors, with higher transaction volumes represented by warmer colors.} 
\label{fig:heatmaps_amount_sector}
\end{figure}

The results obtained by aggregating the transactions with respect to the sector of belonging are shown in Figures~\ref{fig:heatmaps_number_sector} and~\ref{fig:heatmaps_amount_sector}. Note that the sector information was available only for part of the users, amounting to $99\%$, $24\%$, and $34\%$ in 2022, 2023, and 2024, respectively, of the total number of users included in the dataset.


Figure~\ref{fig:heatmaps_number_sector} presents heat maps depicting the volume of transactions across business sectors within the Sardex circuit over the years 2022 (left), 2023 (center), and 2024 (right). In 2022, significant transactional exchanges were observed within the Sardex system. Notably, approximately 50\,k\euro\, were exchanged from the Groceries sector to Employees. Additionally, substantial transactions occurred from the Consumer sector to both Groceries (70\,k\euro) and Wellness (20\,k\euro). Further notable exchanges included 20\,k\euro\, transferred from the Horeca (Hotel, Restaurant, and Catering) sector to Employees, and 12\,k\euro\, from Wellness to Employees.
A similar pattern persisted in 2023, with Groceries continuing to transfer approximately 50\,k\euro\, to Employees. However, a shift in consumer behavior was observed, as the Consumer sector contributed 30\,k\euro\, to Groceries (a decrease from 70\,k\euro\, in 2022) while maintaining a 20\,k\euro\, transaction to Wellness. Additionally, transactions from the Horeca sector to Employees remained stable at 20\,k\euro, as did those from Wellness to Employees, which remained at 12\,k\euro. A more significant shift in transactional exchanges occurred in 2024. The most notable transaction involved a 6\,k\euro\, transfer from Groceries to Horeca. Compared to previous years, the total volume of intersectoral exchanges appeared to decline, suggesting potential structural changes in business interactions within the Sardex ecosystem.

When considering the amount of transactions, the predominant sectoral exchanges in 2022 were led by transactions from Cleaning to Industry and Mechanics, amounting to 40\,k\euro, followed closely by exchanges from Insurance to Music and Events, also totaling 40\,k\euro. In 2023, transactional dynamics shifted, with the highest recorded exchange occurring between Industry and Mechanics and Packaging, reaching 100\,k\euro. Additionally, transactions from Cleaning to Industry and Mechanics increased to 50\,k\euro, while Buildings to Industry and Mechanics recorded a substantial volume of 40\,k\euro. By 2024, transaction volumes exhibited greater diversity but were generally reduced compared to previous years. The most significant exchanges occurred within the Finance and Insurance sector, with intra-sectoral transactions reaching 20\,k\euro, suggesting a shift towards more localized financial interactions and potentially a more balanced distribution of economic exchanges across sectors.


\end{document}